\documentclass[12pt,a4paper]{article}

\usepackage[margin=1in]{geometry}
\usepackage[T1]{fontenc}
\usepackage[utf8]{inputenc}
\usepackage{lmodern}
\usepackage{microtype}
\usepackage{amsmath,amsfonts,amssymb,amsthm,mathtools,bm}
\usepackage{graphicx}
\usepackage{xcolor}
\usepackage{float}
\usepackage[authoryear,round]{natbib}
\usepackage[hidelinks]{hyperref}
\hypersetup{
  pdftitle={A General Framework for Optimal Differentially Private Estimation},
  pdfauthor={Tao Shen, Xin T. Tong, and Wanjie Wang}
}
\graphicspath{{Figure/}}
\newenvironment{subfigure}[2][]{\begin{minipage}[#1]{#2}}{\end{minipage}}

\floatstyle{ruled}
\newfloat{algorithm}{tbp}{loa}
\floatname{algorithm}{Algorithm}
\newcounter{algorithmline}
\newenvironment{algorithmic}[1][1]{%
  \begin{list}{\footnotesize\arabic{algorithmline}:}{%
    \usecounter{algorithmline}%
    \setlength{\leftmargin}{1.8em}%
    \setlength{\labelwidth}{1.3em}%
    \setlength{\labelsep}{0.5em}%
    \setlength{\topsep}{4pt}%
    \setlength{\partopsep}{0pt}%
    \setlength{\itemsep}{2pt}%
    \setlength{\parsep}{0pt}}%
}{\end{list}}
\newcommand{\State}{\item}
\newcommand{\Require}{\item[]\textbf{Require:} }
\newcommand{\Ensure}{\item[]\textbf{Ensure:} }

\theoremstyle{plain}
\newtheorem{thm}{Theorem}
\newtheorem{theorem}[thm]{Theorem}
\newtheorem{pro}{Proposition}
\newtheorem{proposition}[pro]{Proposition}

\theoremstyle{definition}
\newtheorem{defn}{Definition}

\newtheorem{aspt}{Assumption}

\theoremstyle{remark}
\newtheorem{remark}{Remark}

\theoremstyle{plain}
\newtheorem{suppthm}{Theorem}[section]
\newtheorem{supptheorem}[suppthm]{Theorem}
\newtheorem{supppro}[suppthm]{Proposition}

\newtheorem{supplem}[suppthm]{Lemma}

\newtheorem{suppcor}[suppthm]{Corollary}

\theoremstyle{definition}
\newtheorem{suppdefn}[suppthm]{Definition}

\newtheorem{suppaspt}[suppthm]{Assumption}

\theoremstyle{remark}

\definecolor{mypink1}{rgb}{0.858, 0.188, 0.478}
\definecolor{mypink2}{RGB}{219, 48, 122}
\definecolor{mypink3}{cmyk}{0, 0.7808, 0.4429, 0.1412}
\definecolor{mygray}{gray}{0.6}

\definecolor{darkgreen}{rgb}{0,0.4,0}

\newcommand{\myempty}[1]{}

\newcommand{\calB}{\mathcal{B}}
\newcommand{\calN}{\mathcal{N}}

\newcommand{\E}{\mathbb{E}}

\newcommand{\mhat}{\hat{m}}

\newcommand{\thetatilde}{\tilde{\theta}}

\newcommand{\calX}{\mathcal{X}}

\DeclareMathOperator*{\argmax}{arg\,max}

\def\reals{\mathbb{R}}

\newcommand{\ic}{^\top}
\newcommand{\thetahat}{\hat{\theta}}

\newcommand{\fhat}{{\hat{f}}}

\title{A General Framework for Optimal\\
Differentially Private Estimation}
\author{%
  Tao Shen$^{1}$ \qquad Xin T. Tong$^{2}$ \qquad Wanjie Wang$^{1,2}$\\[0.6em]
  {\normalsize $^{1}$Department of Statistics and Data Science}\\
  {\normalsize $^{2}$Department of Mathematics}\\[0.2em]
  {\normalsize National University of Singapore}}
\date{}

\begin{document}
\maketitle

\begin{abstract}
Differential privacy (DP) requires privacy guarantees uniformly over all datasets, whereas statistical guarantees often hold only for typical datasets. We introduce Score-Test-Release (STR), a \emph{simple} and \emph{general} framework that bridges this gap by leveraging high-probability statistical results to construct differentially private estimators. The key is a computationally tractable safety score, constructed using standard perturbation and concentration analyses, that allows STR to exploit the small local sensitivity of typical datasets while maintaining worst-case privacy guarantees. Based on this score, STR privately tests whether to release a Gaussian-perturbed estimate or a data-independent response. We propose a template to find this score. This template allows us to simply develop STR procedures for Bayes classification, linear regression, and kernel regression that achieve minimax-optimal estimation rates up to logarithmic factors under regular assumptions. Numerical studies demonstrate improved accuracy over commonly used DP baselines under comparable privacy guarantees.
\end{abstract}
\noindent\textbf{Keywords:} Differential privacy, Local sensitivity, Minimax optimality, Privacy cost, Score-Test-Release.

\section{Introduction}
Modern statistical analyses often rely on sensitive individual-level data, requiring privacy protection. 
Differential privacy (DP), introduced by \cite{dwork2006differential}, provides a rigorous privacy framework that is applied in many fields \citep{abowd2018us, erlingsson2014rappor, abadi2016deep, bu2020deep}. 
Let $\mathcal X = \{z_1, z_2, \cdots, z_n\}$ and $\mathcal X' = \{z_1', z_2', \cdots, z_n'\}$ be two datasets. 
DP evaluates an individual's privacy protection under neighboring datasets, where $z_i=z'_i$ for all $i\in[n]$ except one index $j$, i.e., the Hamming distance $D_H(\calX, \calX') = 1$.
If it is impossible to identify which dataset the estimate $\tilde\theta$ comes from, then the privacy of $z_j$ is protected. 
Formally, a mechanism $\thetatilde$ is called \emph{$(\varepsilon,\delta)$-DP} if for any measurable set $\calB$ and any datasets $\calX$ and $\calX'$ with $D_H(\calX, \calX') = 1$,
\begin{equation}
\label{eqn:DPformal}
P(\thetatilde(\calX)\in \calB)
\leq
P(\thetatilde(\calX')\in \calB)e^{\varepsilon}+\delta.
\end{equation}
Here $\varepsilon$ and $\delta$ are called the privacy parameters. They evaluate the privacy protection level of the mechanism $\tilde\theta$. The randomness in this definition comes from the design of $\tilde\theta$, not the datasets. 

The randomization in $\tilde\theta$ provides better privacy protection and, meanwhile, induces an accuracy loss in estimation. It naturally brings a question: 
given the privacy parameters $(\varepsilon, \delta)$, what is the optimal estimator? The answer has several parts: the theoretically optimal convergence rate, the computationally tractable optimal rate, and a concrete algorithm that achieves this rate under privacy constraints.  
Generic noise-injection mechanisms are often conservative, leading to large privacy budgets and degraded accuracy. For example, typical choices of $\varepsilon$ range from 1 to 10 in stochastic gradient descent \citep{dong2022gaussian} and large language models \citep{mcmahan2017learning}. 
Recent studies have established differentially private minimax-optimal rates and corresponding procedures in mean estimation and linear regression \citep{cai2021cost}, classification \citep{auddy2025minimax}, and social network analysis \citep{wang2026net}. 
On the other hand, these procedures are often technically involved and designed on a case-by-case basis; a simple, unifying framework is still lacking.

One fundamental difficulty is that statistical accuracy is typically assessed under a data-generating distribution, whereas differential privacy requires guarantees that hold uniformly over all neighboring datasets. Statistical guarantees for an estimator $\hat\theta$ are often established in expectation or with high probability, allowing for irregular behavior on atypical datasets. Differential privacy, in contrast, requires \eqref{eqn:DPformal} to hold for every pair of neighboring datasets $\calX$ and $\calX'$, including those on which $\hat{\theta}$ may be highly unstable. This distinction is particularly consequential for the Gaussian mechanism \citep{dwork2014algorithmic}, which constructs a private estimator $\tilde\theta=\hat\theta+\sigma Z$, where $Z\sim N(0,I_p)$ and $\sigma$ is calibrated to the global sensitivity
$\Delta_{\hat\theta}=\sup_{\calX,\calX':D_H(\calX,\calX')=1}
\|\hat\theta(\calX)-\hat\theta(\calX')\|$.
Although concentration inequalities and random matrix theory may establish that $\hat\theta$ is stable on typical datasets, such high-probability results do not necessarily control $\Delta_{\hat\theta}$, which can be large or even unbounded. Consequently, statistically efficient estimators do not automatically translate into accurate differentially private procedures. Bridging this gap requires mechanisms that exploit statistical stability on typical datasets while preserving worst-case privacy guarantees.

The propose-test-release (PTR) framework exploits local sensitivity while maintaining worst-case differential privacy guarantees \citep{dwork2009differential}. 
The local sensitivity for dataset $\calX$ is the largest change of the estimate $\hat{\theta}(\calX)$ on datasets $\calX'$ with $D_H(\calX, \calX') = 1$. 
PTR first proposes a local sensitivity level $\alpha$, and defines a region $sub_{\alpha}$, on which each dataset has a local sensitivity level at most  $\alpha$. Then, PTR uses a private test to determine whether the observed dataset $\calX$ is sufficiently far from the unstable region $sub_{\alpha}^c$, complement of $sub_{\alpha}$. A noisy estimate is released only when the test is passed; otherwise, the procedure abstains. The safe region $sub_{\alpha}$ avoids the unstable datasets. 
However, although statistical analysis often provides high-probability bounds on local sensitivity, the distance to $sub_{\alpha}^c$ is generally difficult to characterize and compute, even for standard statistical estimators.

To address this computational difficulty, we develop Score-Test-Release (STR), a \emph{simple} and \emph{general} framework for constructing differentially private procedures from existing statistical estimators. The central ingredient is a computationally tractable safety score $\gamma(\calX)=\gamma_{\hat\theta}(\calX;\alpha)$, where $\alpha$ is a proposed local sensitivity level. The score is non-negative and 1-Lipschitz with respect to the Hamming distance, and $\gamma(\calX)>0$ certifies that the local sensitivity of $\hat\theta$ at $\calX$ does not exceed $\alpha$. 
These properties make $\gamma$ a lower bound on the distance to the unstable region, without requiring the exact distance to be computed. Based on this score, STR uses a private test to determine whether to release a Gaussian-perturbed estimate with noise calibrated to $\alpha$, or return a data-independent ``no reply'' message. The release probability increases with the safety score $\gamma(\calX)$, allowing the procedure to exploit local sensitivity while maintaining worst-case differential privacy guarantees.

The key feature of STR is that the safety score $\gamma(\calX)$  can be constructed using familiar tools from statistical analysis. We provide a systematic construction in three steps. First, a deterministic perturbation analysis of $\hat\theta$ identifies a data-dependent regularity statistic $\beta(\calX)$ that controls its local sensitivity. Second, guided by concentration results for $\beta(\calX)$, we choose a threshold $b$ and a sensitivity level $\alpha$ such that $\beta(\calX)\geq b$ implies that the local sensitivity is at most $\alpha$, while $\beta(\calX)$ typically exceeds $b$ under the data-generating model. Finally, if $\beta$ is $L_\beta$-Lipschitz with respect to the Hamming distance, we construct the score 
\[
\gamma(\calX)=\frac{(\beta(\calX)-b)_+}{L_\beta}.
\]
By construction, $\gamma$ is 1-Lipschitz and satisfies the required sensitivity guarantee, ensuring differential privacy through STR. Meanwhile, Steps 1 and 2 leverage perturbation and concentration analyses of $\hat\theta$ to select an appropriate sensitivity level for typical datasets. Together, STR offers a computationally tractable framework to privatize existing estimators while potentially preserving their optimal rates.

We develop the concrete algorithms and theoretical guarantees of STR in three classical statistical problems: Bayes classification, linear regression, and kernel regression. Our main contributions are summarized as follows.
\begin{enumerate}
\renewcommand{\labelenumi}{(\alph{enumi})}
    \item We develop STR, a \emph{general} framework for adapting existing statistical estimators to differential privacy, together with a systematic approach to constructing computationally tractable safety scores $\gamma$. Despite its generality, the construction is \emph{simple}: it uses standard statistical tools and typically takes one to two pages to derive. It thus provides a compact and direct pipeline from a statistical estimator to its DP counterparts.
    \item We establish the theoretical guarantee of STR, including the $(\varepsilon, \delta)$-DP guarantee without distributional assumptions, and a general risk bound under a probabilistic model. 
    The risk bound separates the original estimation error, the additional privacy noise, and the loss due to unsuccessful releases, which deepens our understanding of the risk. 
    \item We apply STR to Bayes classification, linear regression, and kernel regression. For each problem, we construct a simple safety score $\gamma$ and establish statistical error bounds. Under common probabilistic models, the resulting private estimators attain the corresponding minimax rates up to logarithmic factors. 
    These applications demonstrate the flexibility of STR in connecting well-known statistical results and DP. 
    \item We evaluate STR through simulations and real-data analyses across the three applications. The results demonstrate improved statistical accuracy compared with the considered private estimation methods.
\end{enumerate}

Several existing approaches exploit local sensitivity to improve the accuracy of differentially private estimation. Smooth sensitivity calibrates noise using a smoothly varying upper bound on local sensitivity \citep{nissim2007smooth}, with applications to mean estimation \citep{bun2019average}, PCA \citep{gonem2018smooth}, and Naive Bayes classification \citep{zafarani2020differentially}. However, computing such bounds can be challenging for some estimators \citep{alabi2020differentially}. Following the PTR line, \citet{brunel2020propose} and \citet{liu2022differential} develop procedures for mean/median and high-dimensional estimation, respectively, while \citet{wang2022renyi} establishes R\'{e}nyi differential privacy guarantees for PTR.

A related stability-based approach by \citet{guhathakurta2013differentially} constructs computable lower bounds on the distance to instability for private feature selection. But their notion of stability requires the output to remain unchanged under data perturbations. It works for discrete-valued estimators, but is restrictive for continuous estimators such as linear regression and kernel regression. More recently, \citet{redberg2023generalized} generalize PTR to randomized mechanisms. It proposes a data-dependent privacy bound $\epsilon(\calX)$ to summarize randomized output given the input as $\calX$, and then focus on the local sensitivity related to this $\epsilon(\calX)$, utilising PTR.  However, analyzing the local sensitivity again requires an upper bound on local sensitivity, which still faces the same challenges in PTR. In contrast, STR focuses on providing a general template for constructing computationally tractable safety scores directly from the statistical analysis of existing estimators. This template accommodates different statistical problems through their respective perturbation and concentration analyses, as demonstrated in our three applications.

\subsection{Roadmap and notations}
\label{sec:notation}

We introduce the STR procedure with theoretical guarantee in Section \ref{sec:ePTR}, and then demonstrate its effectiveness on three basic statistical problems, Bayes classifier in Section~\ref{sec:bayes}, linear regression in Section~\ref{sec:LR} and kernel regression in Section~\ref{sec:KR}. Proofs and additional numerical
results are provided in the supplementary material.

Throughout the paper, these notations are used. For two
datasets $\mathcal X=\{z_1,\ldots,z_n\}$ and
$\mathcal X'=\{z_1',\ldots,z_n'\}$, define
$D_H(\mathcal X,\mathcal X')=\sum_{i=1}^n \mathbf 1\{z_i\neq z_i'\}$.
Thus, neighboring datasets satisfy \(D_H(\mathcal X,\mathcal X')=1\), denoted by $\calX \sim \calX'$. 
Let $[n]=\{1,\ldots,n\}$ for any positive integer $n$. For a vector $x$, $\|x\|$ denotes the Euclidean
norm; for a matrix $A$, $\|A\|$ denotes the spectral norm. 
For positive sequences $a_n,b_n$, $a_n=O(b_n)$ means
$a_n\le Cb_n$ for some constant \(C>0\) and all sufficiently large \(n\), and $a_n=o(b_n)$
means \(a_n/b_n\to0\). For random sequences, \(a_n=O_p(b_n)\) means \(a_n/b_n\) is bounded
in probability, and \(a_n=o_p(b_n)\) means \(a_n/b_n\to0\) in probability.
For \(R>0\), let \(\Pi_R\) denote projection onto a bounded range. For
a scalar \(x\in\mathbb R\),
$\Pi_R(x)=(-R)\vee(x\wedge R)$.
For a vector \(x\in\mathbb R^p\),
$\Pi_R(x)=\frac{Rx}{\|x\|\vee R}$.

\section{The Score–Test–Release framework}
\label{sec:ePTR}

Let \(\hat{\theta}(\mathcal X)\in\mathbb R^p\) be an estimator to be released privately, where \(p\) denotes the dimension of the released quantity. We first define the safety score and formulate STR, then establish its privacy and statistical guarantees. In Section \ref{subsec:beta}, we show how to construct suitable scores from the regularity properties of \(\hat{\theta}\).

\subsection{The STR framework}
To construct an $(\varepsilon, \delta)$-DP estimator $\tilde\theta$, the sensitivity analysis on the original $\hat{\theta}$ plays the key role. Let $\Delta_{\hat{\theta}} = \sup_{\calX,\calX':\calX'\sim\calX}\|\hat{\theta}(\calX) - \hat{\theta}(\calX')\|$ denote the global sensitivity. Then the Gaussian mechanism designs a private estimator that 
\[
\tilde\theta_{global} = \hat{\theta} + \frac{\sqrt{2\log(1.25/\delta)}}{\varepsilon} \Delta_{\hat{\theta}} Z, \qquad Z \sim \mathcal{N}(0, I_p). 
\]
While $\tilde\theta_{global}$ is $(\varepsilon, \delta)$-DP, $\Delta_{\hat{\theta}}$ can be very large or even infinite because of the instability in rare datasets.

To incorporate high-probability analysis results, we focus on local sensitivity, which captures the instability of a given dataset. 
Denote the local sensitivity of $\hat{\theta}$ on a dataset $\calX$ as 
\begin{equation}\label{eqn:localsensitivity}
\Delta_{\hat{\theta}}(\mathcal X)
=
\sup_{\calX':\mathcal X'\sim\mathcal X}
\|
\hat{\theta}(\mathcal X)-\hat{\theta}(\mathcal X')
\|.
\end{equation}
It measures the largest possible change in $\hat{\theta}(\calX)$ if we replace a data point in $\calX$. 
If all data points fall into the high-probability bulk, replacing one data point usually cannot induce a rapid change in the estimator $\hat{\theta}$. 

Based on local sensitivity, PTR assesses the stability of a dataset and privately determines whether to release a privatized estimate. For a proposed sensitivity level $\alpha>0$, define the stable region
\begin{equation}\label{eqn:stableregion}
sub_{\alpha} = sub_{\alpha}(\Delta_{\hat{\theta}})
=
\left\{
\mathcal X:
\Delta_{\hat{\theta}}(\mathcal X)\leq\alpha
\right\}.
\end{equation}
The private test in PTR is based on the distance from the observed dataset $\calX$ to the unstable region $sub_{\alpha}^c$, the complement of $sub_{\alpha}$. 
Let $D_H(\calX, sub_{\alpha}^c) = \inf_{\calX' \in sub_{\alpha}^c} D_H(\calX, \calX')$. 
PTR releases a noisy estimate $\hat{\theta}(\calX)+(\sqrt{2\log(1.25/\delta)}\alpha/\varepsilon)Z$, where $Z\sim N(0,I_p)$, with a probability increasing in $D_H(\calX, sub_{\alpha}^c)$; otherwise, it returns a data-independent ``no reply'' message $\bot$. However, PTR can be computationally intractable: the unstable region $sub_{\alpha}^c$ rarely has a tractable description, and computing the Hamming distance $D_H(\calX, sub_{\alpha}^c)$ is in general a combinatorial optimization problem.

We introduce a safety score $\gamma(\calX)$ to replace the exact distance $D_H(\calX,sub_{\alpha}^c)$.
\begin{defn}(Safety Score)
Given an estimator \(\hat{\theta}\) and a sensitivity level \(\alpha>0\), a non-negative function \(\gamma=\gamma_{\hat{\theta}}(\,\cdot\,;\alpha)\) is called an \(\alpha\)-safety score if, for all datasets \(\mathcal X,\mathcal X'\),
\[
\left|
\gamma(\mathcal X)-\gamma(\mathcal X')
\right|
\leq
D_H(\mathcal X,\mathcal X'),
\qquad 
\left\{\mathcal X:\gamma(\mathcal X)>0\right\}
\subseteq {sub}_{\alpha}(\Delta_{\hat{\theta}}).
\]    
\end{defn}

The first condition ensures that the score $\gamma$ is stable to data perturbations. The second ensures that a positive score certifies local sensitivity at most \(\alpha\). Putting them together indicates that $\gamma$ evaluates how far away $\calX$ is from the unstable region. 
An example is $\gamma(\calX) = D_H(\calX, sub^c_{\alpha})$, the exact distance to $sub_{\alpha}^c$. 
More tractable choices of $\gamma$ can instead substantially reduce the computational cost. 

For a specified \(\alpha\)-safety score, STR uses a randomized test to determine whether to release a Gaussian-perturbed estimate. The details are in Algorithm \ref{alg:ePTR}.

\begin{algorithm}[!ht]
\caption{Score–Test–Release (STR)}\label{alg:ePTR}
\begin{algorithmic}[1]
\Require Data $\calX$, privacy parameter $0 < \varepsilon < 2$ and $0 < \delta < 1$, local sensitivity level $\alpha > 0$, and an $\alpha$-safety score $\gamma$.
\Ensure An $(\varepsilon,\delta)$-DP output $\thetatilde_\gamma(\calX)$.
\State Let
$M=1+\frac{2}{\varepsilon}
\log(
\max\left\{\delta^{-1}, \varepsilon^{-1}\right\}
)$.

\State {\bf (Score)}. Compute $\gamma(\calX)$ 
\State {\bf (Test)}. Set
$$
p(\calX)
=
\frac{
\exp\left(\frac12\varepsilon\{\gamma(\calX)-M\}\right)
}{
\exp\left(\frac12\varepsilon\{\gamma(\calX)-M\}\right)+1
}.
$$
Draw $I\sim\mbox{Bernoulli}(p(\calX))$.
\State {\bf (Release)}. 
If $I = 1$, set
\(    \thetatilde_{\gamma}
    =
    \thetahat(\calX)
    +
    \frac{2\alpha}{\varepsilon}
    \sqrt{2\log(1.25/\delta)}\,\zeta\),
    \(\zeta\sim\calN(0,I_p)\). Otherwise, set $\thetatilde_{\gamma}=\bot$, a ``no reply'' message.
\end{algorithmic}
\end{algorithm}

\begin{remark}
The ``no reply'' message $\bot$ represents a data-independent output in the parameter space. It may be fixed or drawn independently from a prespecified distribution $\pi_{\bot}$. 
\end{remark}

The release probability increases with the safety score and equals $1/2$ when \(\gamma(\mathcal X)=M\). Thus, a positive score certifies the sensitivity bound, while a larger score makes release more likely. We next establish the privacy guarantee of STR with any valid $\alpha$-safety score $\gamma$. The statistical performance of STR under a probabilistic model is discussed in Section \ref{subsec:eptr_performance}, illustrating the role of private tests and $\gamma$. We end this section with a template of choosing $\gamma$ utilizing statistical analysis. 

\subsection{Privacy guarantee: STR is DP}
The privacy guarantee can be established under the definition of $\alpha$-safety score, without any assumptions on the data-generating distribution. 
\begin{theorem}
\label{thm:ePTRissafe}
    Let \(0<\varepsilon<2\), \(0<\delta<1\), and \(\alpha>0\). If \(\gamma\) is an \(\alpha\)-safety score for \(\hat{\theta}\), then the output \(\tilde{\theta}_{\gamma}\) of Algorithm \ref{alg:ePTR} is \((\varepsilon,\delta)\)-differentially private.
\end{theorem} 
\begin{proof}[Proof sketch] 
Consider $\calX \sim \calX'$, and let $I$ denote the release indicator. Since $\gamma$ is 1-Lipschitz, together with the logistic release rule, there is 
$P(I(\calX)=i)\leq e^{\varepsilon/2}P(I(\calX')=i)$
for $i\in\{0,1\}$. Thus, the private test is $(\varepsilon/2,0)$-DP.

If at least one of $\gamma(\calX)$ and $\gamma(\calX')$ is positive, the sensitivity guarantee of the safety score implies
$\|\hat\theta(\calX)-\hat\theta(\calX')\|\leq\alpha$.
The calibrated Gaussian release therefore satisfies the
$(\varepsilon/2,\delta)$ privacy inequality for this pair.
Hence, the estimator is $(\varepsilon, \delta)$-DP.

If both scores are zero, the common release probability is
$(1+e^{\varepsilon M/2})^{-1}\leq\delta$ by the choice of $M$.
Since the non-release distribution is identical for both datasets,
their output probabilities differ by at most $\delta$.
The same privacy inequality therefore holds in this case,
completing the proof.
\end{proof}

\subsection{Statistical accuracy: the performance of STR}\label{subsec:eptr_performance}
We next study the statistical performance of STR estimator $\tilde\theta$ under a probabilistic model $P_{\theta}$. We measure accuracy by mean squared error
$\mathbb E_{\theta}
\bigl[
\|
\tilde{\theta}_{\gamma}(\mathcal X)-\theta
\|^2
\bigr]$, which includes both the data randomness and the privacy mechanism randomness. The following theorem establishes a risk bound of $\tilde\theta$. 
\begin{thm}
\label{thm:ePTRperformance}
Given any $P_{\theta}$ parameterized by $\theta\in \reals^p$, 
suppose the no reply message $\|\bot-\theta\|^2\leq C_\bot$ almost surely for some constant $C_\bot$. Then for STR estimate $\thetatilde$, 
\begin{equation*}
\E_{\theta}[\|\thetatilde(\calX)-\theta\|^2]\leq \underbrace{\E_\theta [\|\thetahat(\calX)-\theta\|^2]}_{\text{(I)}}+\underbrace{\frac{8\alpha^2p}{\varepsilon^2}\log(\frac{1.25}{\delta})}_{\text{(II)}}+\underbrace{C_\bot(P_\theta(\gamma(\calX)<2M)+\delta)}_{\text{(III)}}.
\end{equation*}
\end{thm}
The error can be decomposed into three terms. Term (I) is the privacy-less error. It is the estimation error of the original estimator $\hat{\theta}$ and is unrelated to the privacy mechanism. To control it, we use standard statistical literature results. Term (II) is the privacy perturbation loss. It comes from the calibrated Gaussian noise. Hence, a tight local sensitivity level $\alpha$ reduces this error. Term (III) is the non-release loss, which accounts for the default output $\bot$ when the test fails. If $\gamma$ is not tight, then there will be higher probability to fail and hence Term (III) is larger.

To illustrate terms (II) and (III), we consider simple linear regression as an example. Suppose the dataset $\calX = \{(x_1, y_1), \cdots, (x_n, y_n)\}$, where $y_i=\beta x_i+\xi_i$, $x_i\sim N(0,1)$ and $\xi_i\sim N(0,\sigma^2)$ are independent. The ordinary least square (OLS) estimator is $\hat\beta=\sum_i x_iy_i/\sum_i x_i^2$. After clipping the observations and bounding the estimator, the denominator $\sum_i x_i^2 \asymp n$ with high probability, yielding local sensitivity of order $\log n/n$. 
It suggests choosing $\alpha\asymp\log n/n$ and $\delta\asymp n^{-3}$, and Term (II) is at order $\log^3 n/(n^2\varepsilon^2)$. Term (III) evaluates the failure probability. With a suitable choice of constants in $\alpha$, taking $\gamma = D_H(\calX, sub_{\alpha}^c)$ leads to a probability that $P_{\theta}(\gamma(\calX) < 2M) \leq  1/n$. Hence, Term (III) is of order $n^{-1}$. Thus, both terms can be controlled at a similar rate as the non-private OLS convergence rate. This toy example illustrates the trade-off between (II) and (III), and the preference on $\alpha$. 

The risk decomposition is similar to that for the standard PTR in \cite{brunel2020propose}. Our goal is to achieve the same performance using a computationally tractable safety score.

\subsection{Constructing safety scores}
\label{subsec:beta}
We provide a general three-step template for constructing a computationally tractable safety score. Given an existing estimator $\hat\theta$, the construction centres on three quantities: a data-dependent regularity statistic $\beta(\calX)$ based on $\hat\theta$, the local sensitivity level $\alpha$, and the resulting safety score $\gamma$. 

\begin{enumerate}
\renewcommand{\labelenumi}{(\alph{enumi})}
    \item \textbf{Perturbation analysis}. Analyze the local sensitivity $\Delta_{\hat\theta}(\calX)$ and identify a computable statistic $\beta(\mathcal X)$. This $\beta(\calX)$ captures the fluctuation so that for any small $\alpha$, there is a constant $b$, so that for every dataset $\calX$, 
\[
\beta(\mathcal X)\geq b
\quad\Longrightarrow\quad
\Delta_{\hat\theta}(\mathcal X)\leq\alpha.
\]

\item \textbf{Concentration analysis}. Study the concentration of $\beta(\mathcal X)$ under the probabilistic model. The concentration analysis guides the choice of $b$ and hence $\alpha$. Ideally, we would like to keep typical datasets, which means $\beta(\mathcal X) \geq b$ with high probability, while $\alpha$ is as small as possible.

\item \textbf{Safety score construction}. If $\beta$ is $L_\beta$-Lipschitz with respect to Hamming distance, define
\[
\gamma(\mathcal X)=\frac{(\beta(\mathcal X)-b)_+}{L_\beta},
\]
where $(x)_+=\max\{x,0\}$. Then this score $\gamma$ is a valid $\alpha$-safety score, and it can be used to implement STR. The proof is easy and omitted here. 
\begin{proposition}
Suppose that $\beta$ is $L_\beta$-Lipschitz with respect to Hamming distance, and $\beta(\calX) \geq b$ indicates that $\Delta_{\hat{\theta}}(\calX) \leq \alpha$.
Then $\gamma(\mathcal X)=(\beta(\mathcal X)-b)_+/L_\beta$ is an $\alpha$-safety score for $\hat\theta$. 
\end{proposition}
\end{enumerate}
These three steps bridge high-probability statistical results and the construction of  differentially private estimators. 
Perturbation and concentration analyses establish that typical datasets have small local sensitivity, which STR can exploit to control the perturbation error in Term (II). For datasets where this sensitivity bound cannot be certified, STR allows a data-independent non-release response. Since typical datasets pass the private test with high probability, Term (III) also remains small. 
By Theorem \ref{thm:ePTRperformance}, STR thus leverages the statistical properties of typical datasets while guaranteeing privacy for all datasets, providing a route to rate-optimal differentially private estimation.

Sections \ref{sec:bayes}–\ref{sec:KR} implement this construction for Bayes classification, linear regression and kernel regression. In these applications, the regularity statistics arise from class sample sizes, the smallest eigenvalue of the design matrix, and the local kernel weight, respectively. We derive explicit safety scores with this template, establish the resulting risk bounds, and demonstrate the optimality of the final STR estimator. These illustrate the generality of the STR template.

\section{Bayes classifier}\label{sec:bayes}
We start with a basic statistical problem --- classification \citep{bayes}. The Bayes classifier is one of the most basic classification tools. 
Consider a simple setup where the class labels $y_i$ come from a prior $P(y_i=k)=\mu_k$, and the covariates follow Gaussian $x_i|y_i=k \sim \mathcal{N}(m_k,I)$, $x_i, m_k \in \reals^p$. 
With known parameters, for a new data point $x$, the naive Bayes classifier considers the posterior probability and classifies it into the class with maximum posterior probability, i.e., 
\begin{equation}\label{eqn:bayes}
\hat{y}(x) = \argmax_{1 \leq k \leq K} \frac{\exp(-\frac12 \|x-m_k\|^2)\mu_k}{\sum_{j\in[K]} \mu_j\exp(-\frac12 \|x-m_j\|^2)} = \argmax_{1 \leq k \leq K} \mu_k e^{-\frac12 \|x-m_k\|^2}.
\end{equation}
It is also connected to logistic regression. 

In practice, the parameters are unknown. We estimate the prior $\mu_{[K]}$ using the sample frequency and mean $m_{[K]}$ using the sample mean. Consider the dataset $\calX=\{(x_i,y_i), i\in [n]\}$, we have 
\begin{equation}
\label{eqn:bayesamplemean}
\hat{\mu}_k(\calX)=\frac{|I_k|}{n},\quad \mhat_k(\calX)=\frac{\sum_{j\in I_k} x_j}{|I_k|},\quad \text{where } I_k:=\{j:y_j=k\}.
\end{equation}
The practical naive Bayes classifier follows by introducing $\hat{\mu}_k$ and $\hat{m}_k$ into \eqref{eqn:bayes}. 
In this paper, for an expression $A^{-1}B$, when $A$ is singular, e.g. $|I_k|=0$ in \eqref{eqn:bayesamplemean},  we treat $A^{-1}B$ as zero. 

\subsection{Formulating STR}\label{subsec:eptr_bayes}
We now apply the STR framework in Section~\ref{sec:ePTR} to construct
a private Bayes classifier. Replacing $\mu_k$ and $m_k$ in
\eqref{eqn:bayes} with private estimates $\tilde\mu_k$ and
$\tilde m_k$ requires us to privatize the estimators in
\eqref{eqn:bayesamplemean}. The first step is to find a computationally
tractable statistic $\beta(\calX)$ that controls their local sensitivity.
We therefore examine how the estimators change when one observation is
replaced.

Consider a dataset $\calX=\{(x_i,y_i):i\in[n]\}$. For simplicity of
exposition, assume that $\|x_i\|\le R_x$ for a known $R_x$; this can be
enforced by projecting the observations onto the ball of radius $R_x$.
Let $\calX_i'$ differ from $\calX$ only in its $i$-th observation
$(x_i',y_i')$. The class-proportion estimator $\hat\mu_k=|I_k|/n$ has
global sensitivity at most $1/n$. For the class-mean estimator
$\hat m_k$, provided that $|I_k|\ge 2$, changing $(x_i,y_i)$ gives
\begin{align}
\label{eqn:bayesperturb}
\Delta_{\mhat_k}(\calX)
&=
\max_{\calX'\sim\calX}
\|\mhat_k(\calX)-\mhat_k(\calX')\| \notag\\
&\le
\max\left\{
\frac{R_x+\max_{j\in I_k}\|x_j\|}{|I_k|},
\frac{\max_{j\in I_k}\|x_j-\mhat_k\|}{|I_k|-1},
\frac{\max_{x:\|x\|\le R_x}\|x-\mhat_k\|}{|I_k|+1}
\right\} \notag\\
&\le \frac{2R_x}{|I_k|-1}.
\end{align}
The three terms correspond, respectively, to replacing an observation
within class $k$, removing one from class $k$, and adding one to class
$k$. The bound shows that the smallest class count controls the local
sensitivity across all class means. We accordingly define
\(
\beta(\calX)=\min_{k\in[K]}|I_k|-1.
\)
Whenever $\beta(\calX)>0$, \eqref{eqn:bayesperturb} yields
\(
\max_{k\in[K]}\Delta_{\mhat_k}(\calX)
\le \frac{2R_x}{\beta(\calX)}.
\)
Thus, $\beta$ provides the tractable sensitivity bound required by the
first step of the STR construction.

We next choose a sensitivity threshold $\alpha$ that applies to typical
datasets. Suppose that the class probabilities satisfy
$\min_{k\in[K]}\mu_k\ge\mu_{\min}>0$, and choose a constant
$c_0\in(0,\mu_{\min})$. Then $\beta(\calX)\ge c_0n$ with high probability
as $n$ grows. Setting
\(
\alpha=\frac{2R_x}{c_0n}
\)
therefore ensures that
\[
\max_{k\in[K]}\Delta_{\mhat_k}(\calX)\le\alpha
\qquad\text{whenever}\qquad
\beta(\calX)\ge c_0n.
\]
The restriction to typical datasets matters here: a class may contain
only one observation on some datasets, so the local sensitivity of its
mean does not uniformly decrease at rate $n^{-1}$. The lower bound on
the class probabilities ensures that such small class counts become
unlikely, but it does not exclude them from a differential privacy
guarantee.

Finally, we construct a safety certificate for the chosen $\alpha$.
Replacing one observation changes each class count by at most one and
hence changes $\beta(\calX)$ by at most one. Thus, $\beta$ is
$1$-Lipschitz under the Hamming distance. Following the construction
in Section~\ref{sec:ePTR}, we set
\begin{equation}
\label{eqn:gammaBC}
\gamma(\calX)
=
\bigl(\beta(\calX)-c_0n\bigr)_+
=
\bigl(\min_{k\in[K]}|I_k|-c_0n-1\bigr)_+.
\end{equation}
The function $\gamma$ is $1$-Lipschitz, and
\(
\{\gamma(\calX)>0\}
\subseteq
\left\{
\max_{k\in[K]}\Delta_{\mhat_k}(\calX)\le\alpha
\right\}.
\)
Consequently, $\gamma$ is an $\alpha$-safety score. Both $\alpha$ and
$\gamma(\calX)$ are available in closed form from the class counts.

Now recall that we have the proportion estimate $\hat{\mu}_k$. Its sensitivity follows that 
$|\hat{\mu}_k(\calX)-\hat{\mu}_k(\calX')|\leq 1/n$.    
This overall sensitivity is already small.

With these preparations, we propose an STR adaptation for the Bayes classifier in \eqref{eqn:bayesamplemean}. 
To meet the requirement that $\|x_i\| \leq R_x$, the algorithm involves a projection step
$\Pi_{R}(x)=\frac{Rx}{\|x\|\vee R}$. 
This is a standard step in many DP algorithms, see \cite{cai2021cost}. 
We further revise the local sensitivity $\alpha$, so that releasing both $\tilde{\mu}_{k}$ and $\tilde{m}_{k}$ for all $k$ at the same time is well controlled. 
Details are  in Algorithm \ref{alg:ePTRBC}. 
\begin{algorithm}
\caption{Bayes Classifier with STR (STR-Bayes)}\label{alg:ePTRBC}
\begin{algorithmic}[1]
\Require Data $\calX$, tolerance $\varepsilon,\delta$, prefixed bounds $R_x, c_0$.
\Ensure An $(\varepsilon,\delta)$-DP release $\thetatilde(\calX)$ and classifier $\tilde{y}$.
\State Let $x_i=\Pi_{R_x}(x_i)$. Find $\hat{\mu}_k,\mhat_k.$
\State Apply STR with parameter $\varepsilon, \delta$ on $(\hat{\mu}_{[K]},\mhat_{[K]})$ with $\gamma$ being \eqref{eqn:gammaBC} and $\alpha=\frac{2}{n}\sqrt{\frac{2R^2_x}{c^2_0 }+2}$. Denote $(\tilde{\mu}_{[K]}, \tilde{m}_{[K]})$ as the output. 
\State Revise  $\tilde{\mu}_{[K]}$ to be probability vector
\[
\tilde{\mu}_{k}  \leftarrow \frac{\tilde{\mu}_{k}\vee c_0}{\sum^K_{j=1}\tilde{\mu}_{j}\vee c_0}.
\]
\State Release $\tilde\mu_{[K]}$ and $\tilde{m}_{[K]}$. The classifier follows that 
\[
\tilde{y}(x) = \argmax_{1 \leq k \leq K} (\tilde{\mu}_k e^{-\frac12 \|x-\tilde{m}_k\|^2}).
\]
\end{algorithmic}
\end{algorithm}
\begin{theorem}
\label{thm:BayePTRDP}
The output of Algorithm~\ref{alg:ePTRBC} is $(\varepsilon,\delta)$-DP for $\varepsilon\in (0,2)$. 
\end{theorem}

\subsection{Performance analysis}
In this section, we examine the error bound of STR-Bayes in Algorithm \ref{alg:ePTRBC}. 
Unlike DP guarantee, risk analysis requires some model assumptions. We make standard assumptions as follows. 
\begin{aspt}
\label{aspt:Bayes}
The data points $(x_i,y_i)$ are obtained as i.i.d. samples following 
\[
P(y_i=k)=\mu_k,\quad x_i|{y_i=k}\sim \mathcal{N}(m_k, I).
\]
\end{aspt}
By Theorem \ref{thm:ePTRperformance}, we would like to control the local DP loss indicated by $\alpha$ and the STR failure loss. 
Suppose the $\mu_{[K]}$ satisfies $\min_{k \in [K]} \mu_k>2c_0$, then a Bernstein inequality indicates $\gamma(\calX)>\frac12c_0 n$ with high probability. 
Applying this in Theorem \ref{thm:ePTRperformance}, we see that if $\varepsilon\gg \log(1/\delta)/n$, $\delta<1/n$ then $P_\theta(\gamma(\calX)>2M)=1-o(1)$. So the STR loss is negligible. 
\begin{theorem}
\label{thm:BayePTR}
Under Assumption \ref{aspt:Bayes}, suppose $\min_{k\in[K]} \mu_k>2c_0$, and there exists a constant $C_m \geq 1$ so that $\|m_k\|\leq C_m\sqrt{p}$. Let $\tilde\mu_{[K]}$ and $\tilde{m}_{[K]}$ be the STR-Bayes output. Taking
$R_x=4C_m\sqrt{p|\log n|}$ and $\bot=0$, $\delta<n^{-3}$, 
and ${8|\log ({\delta})|}/{\varepsilon } \leq c_0n - 4$, then
\[
\E[\sum_{k=1}^K(|\mu_k - \tilde{\mu}_k|^2+\|m_k-\tilde{m}_k\|^2)]
\leq C\left(\frac{p}{n}+\frac{p^2\log n\log (1/\delta)}{\varepsilon^2 n^2}\right).
\]
\end{theorem}
When there is one class $K=1$, the optimal mean estimation loss is $O(\frac{p}{n}+\frac{p^2\log (1/\delta)}{\varepsilon^2 n^2})$ \citep{cai2021cost}. Our method matches this lower bound up to a $\log n$ inflation, which yields near optimality. 

\subsection{Numerical simulation}

We evaluate the proposed STR-Bayes classifier under the Gaussian Bayes model
introduced earlier, with $K=3$ and $p=10$.
Specifically, the training observations are $\{(x_i,y_i): i\in[n]\}$, where
$P(y_i=k)=\mu_k$ and $x_i\mid y_i=k \sim \mathcal N(m_k,I_p)$.
The class means are taken as coordinate-separated signals,
\[
    m_1=(\Delta,0,\ldots,0)^\top,\quad
    m_2=(0,\Delta,0,\ldots,0)^\top,\quad
    m_3=(0,0,\Delta,0,\ldots,0)^\top,
\]
with $\Delta=3$. Throughout, the observed covariates are clipped to the Euclidean ball of radius $R_x=8$ before estimation.

We consider three regimes, varying the privacy budget \(\varepsilon\), sample size \(n\), and minimum class proportion \(\pi_{\min}\). For the first two, we set \((\mu_1,\mu_2,\mu_3)=(0.75,0.15,0.10)\); for the last one, we set \((\mu_1,\mu_2,\mu_3)=(0.7-\pi_{\min},0.3,\pi_{\min})\). For STR-Bayes, we set $c_0=0.04$ in the first two regimes
and $c_0=\pi_{\min}/2$ in the class-imbalance regime. Performance is evaluated by balanced test error, namely the average misclassification rate across classes, on an independent test set of size \(10^5\), averaged over \(500\) replications. Although the overall misclassification rate is a natural measure of classification performance, we use balanced test error to give equal weight to each class, consistent with our theoretical focus on estimation accuracy across classes. This metric is particularly informative under class imbalance, as it prevents the majority class from dominating the evaluation. We compare the non-private Gaussian naive Bayes classifier, STR-Bayes (STR), DP naive Bayes (DPNB) \citep{vaidya2013differentially}, and its smooth-sensitivity variant (DPNBSS) \citep{zafarani2020differentially}. Unless otherwise stated, \(\delta=0.01\). When $\varepsilon>2$, we verify the same Gaussian mechanism is DP using \cite{balle2018improving}.

\begin{figure}[!htbp]
    \centering
    \begin{subfigure}[t]{0.3\textwidth}
        \centering
        \includegraphics[width=\linewidth]{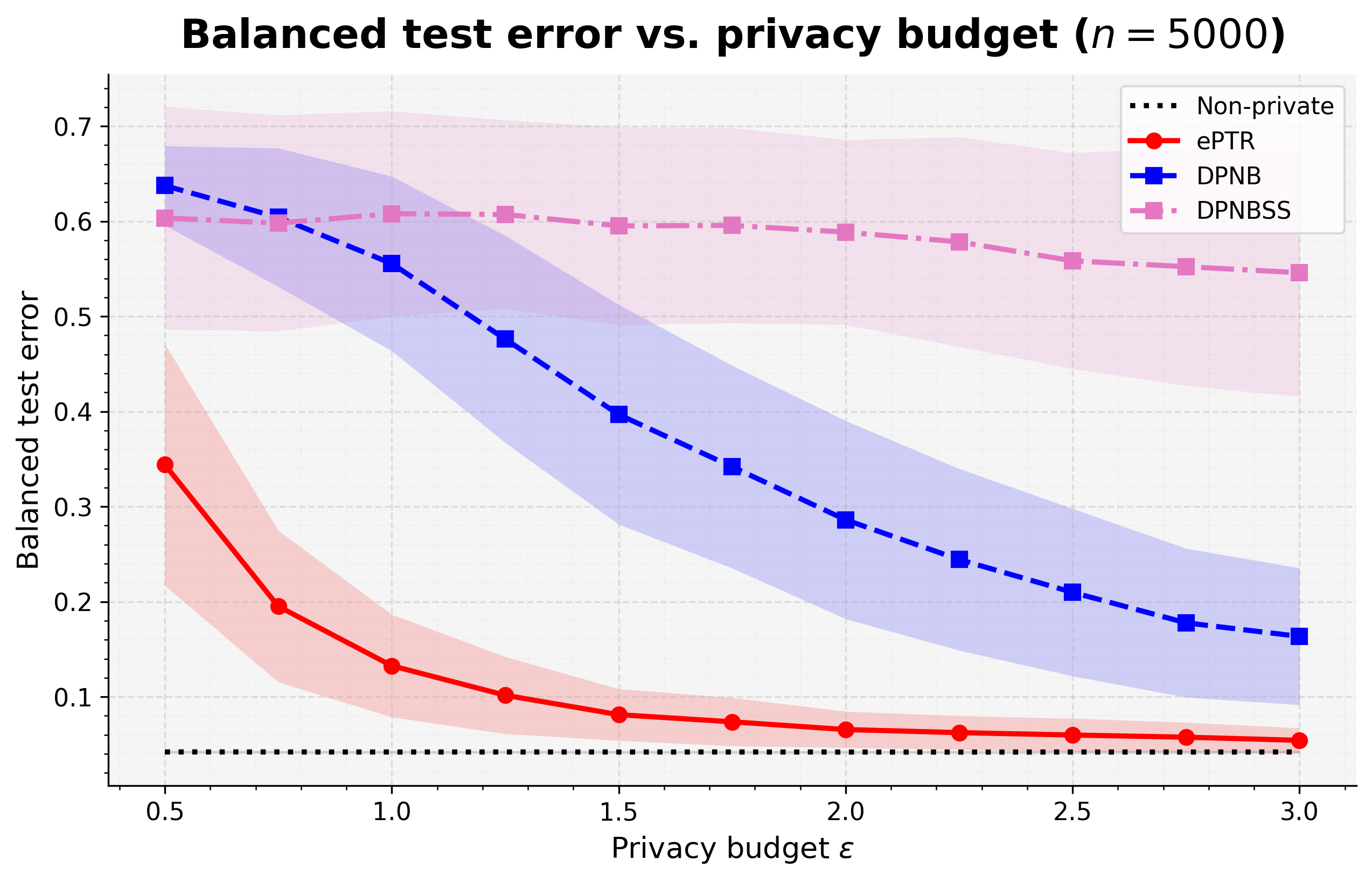}
    \end{subfigure}
    \begin{subfigure}[t]{0.3\textwidth}
        \centering
        \includegraphics[width=\linewidth]{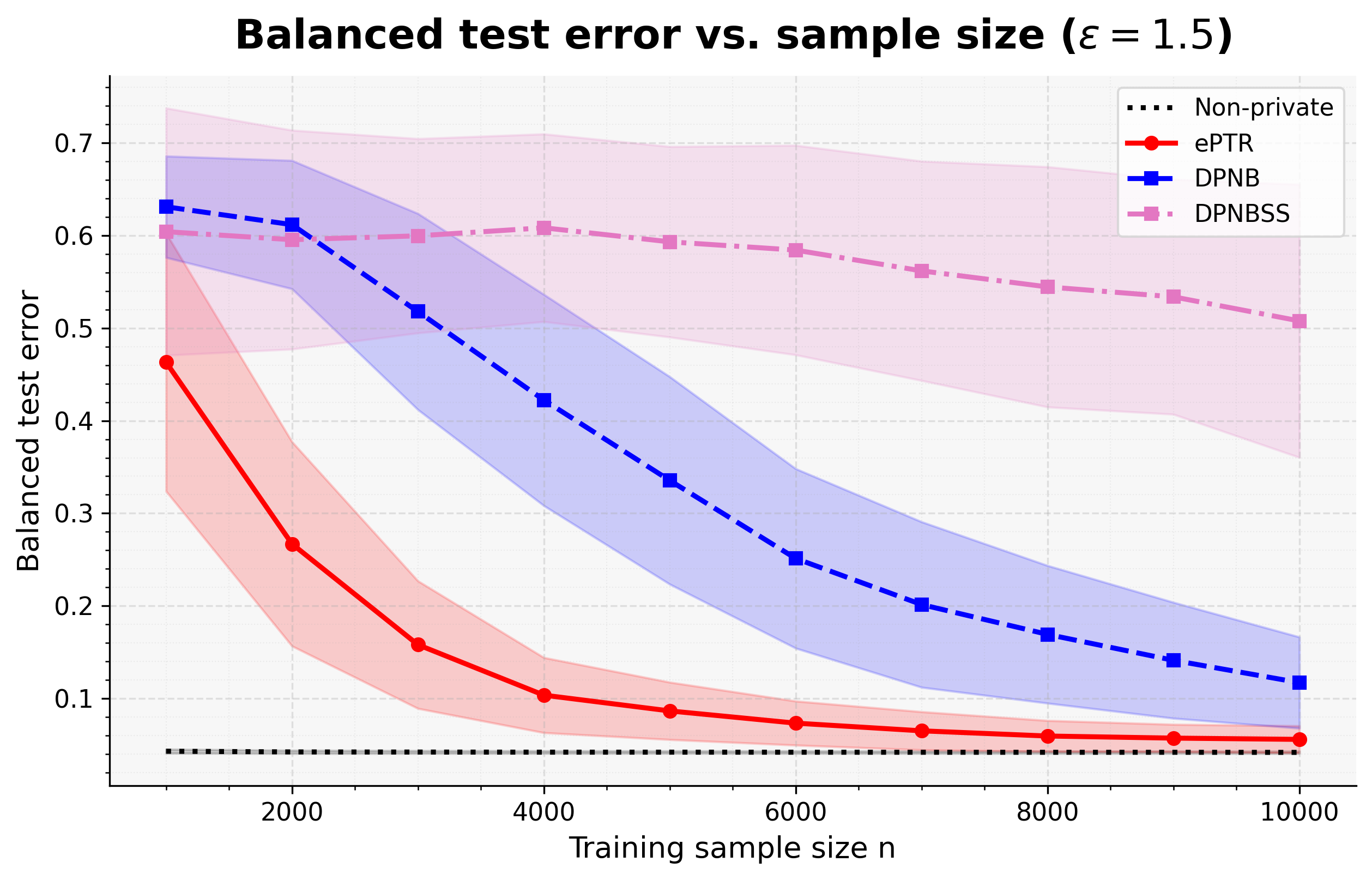}
    \end{subfigure}
    \begin{subfigure}[t]{0.3\textwidth}
        \centering
        \includegraphics[width=\linewidth]{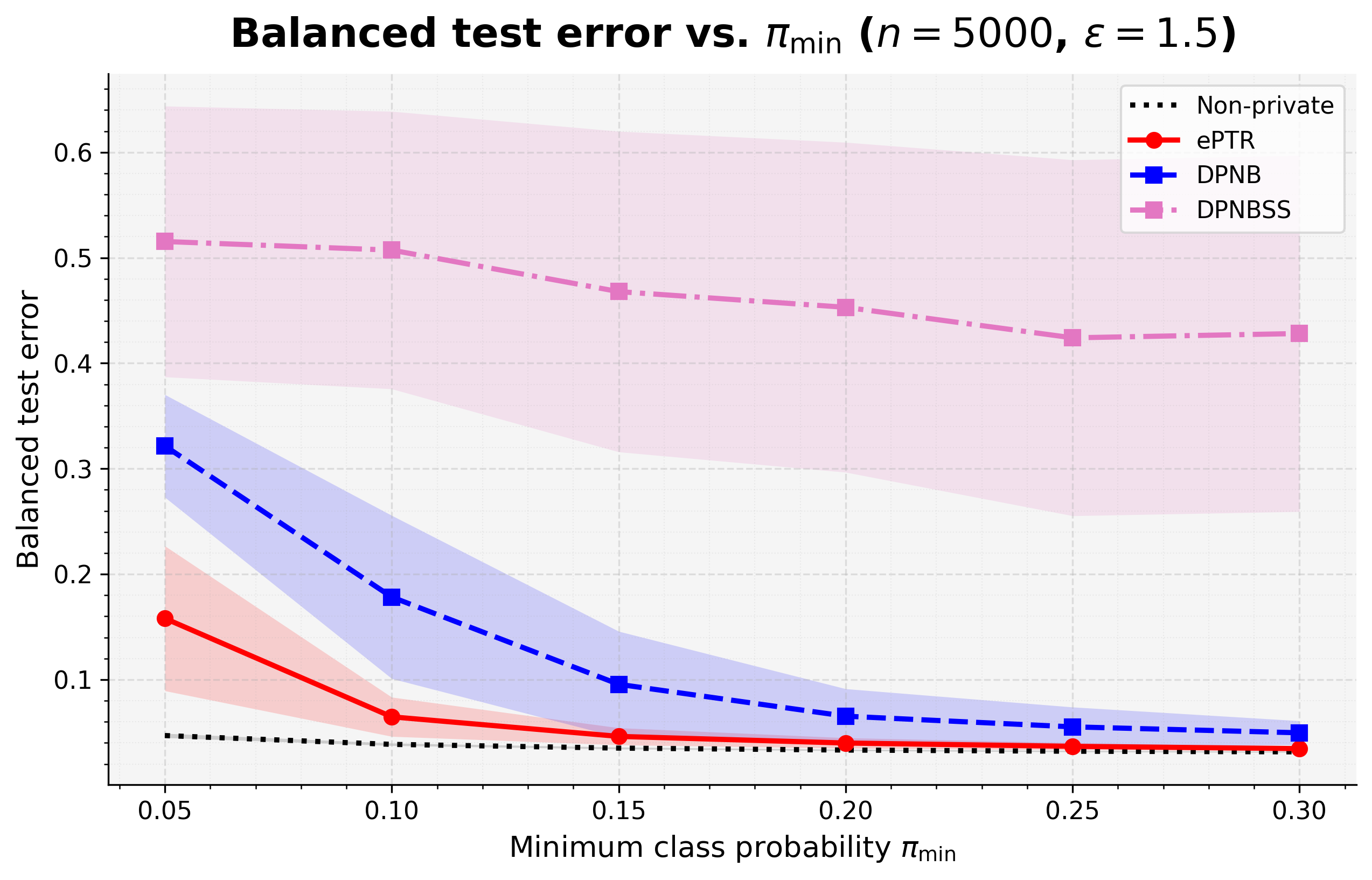}
    \end{subfigure}
    \includegraphics[width=0.90\textwidth]{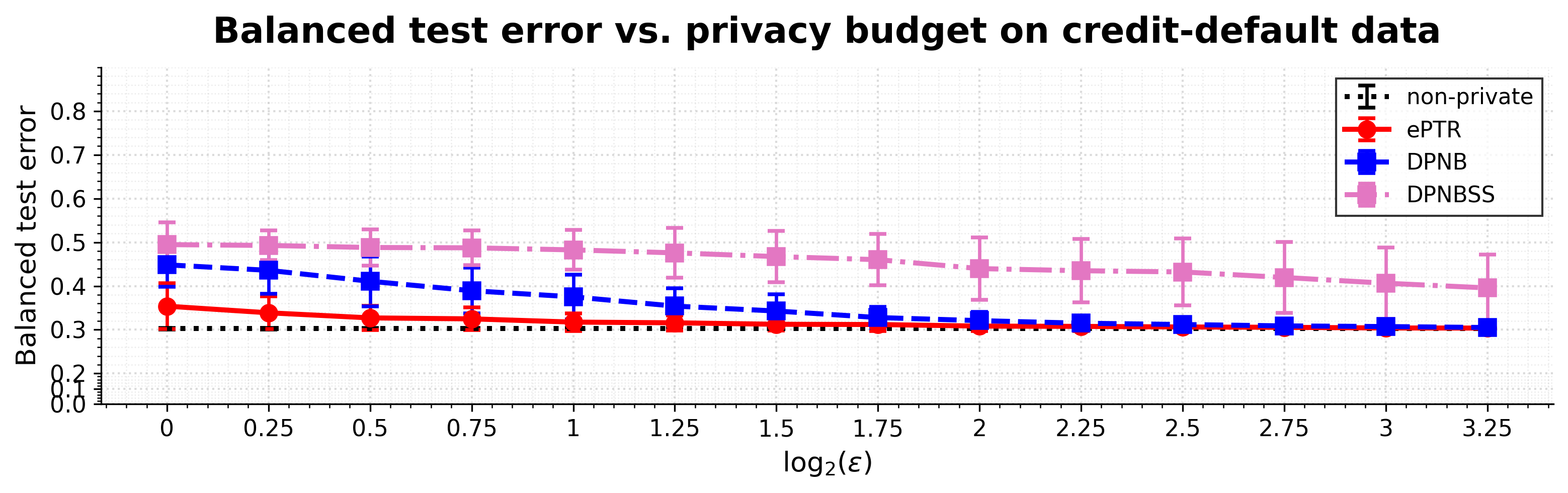}
    \caption{Top row: Simulation results for Bayes classification under varying privacy budget, sample size, and class imbalance condition. Bottom row: balanced test error versus $\log_2(\varepsilon)$ for Bayes classification on credit default data.}
    \label{fig:bayes_sim_all}
    
\end{figure}

Figure~\ref{fig:bayes_sim_all} summarizes the main findings. As expected, the non-private classifier achieves the lowest balanced error throughout. Among the private methods, STR performs most favorably: its error decreases as $\varepsilon$ or $n$ increases, and it consistently attains lower balanced error than DPNB and DPNBSS across all settings. The advantage of STR is especially clear under class imbalance, where rare classes make private estimation more challenging for all methods, but STR still tracks the non-private benchmark much more closely than the competing private alternatives.

\subsection{Credit default classification}
We consider the Default of Credit Card Clients dataset from the UCI Machine Learning Repository, which contains $30{,}000$ credit-card client records from Taiwan \citep{yeh2009comparisons}. We apply
Bayes classification with the default indicator as the binary response. The
covariates include credit limit, demographic variables, repayment history, bill
amounts, and previous payments. Because these variables describe sensitive
individual-level financial behavior, private release is especially natural in
this application. 
As shown in Figure~\ref{fig:bayes_sim_all}, the STR-Bayes classifier with $R_x=15$ and $c_0=0.08$ follows the non-private
Bayes benchmark closely across the privacy-budget range. In contrast, other methods exhibit larger balanced error, especially when
$\log_2(\varepsilon)$ is small. This suggests that STR can
preserve much of the non-private classification accuracy while avoiding unstable private releases.

\section{Linear regression}\label{sec:LR}
In this section, we discuss the multiple linear regression. 
Suppose the observed data points are $(x_i, y_i)$, where $x_i \in \reals^p$ and the model follows
\(y_i=x_i\ic \theta+\xi_i, E[\xi_i] = 0.\)
 Our goal is to estimate the parameter vector $\theta\in \reals^p$. The well-known ordinary least squares (OLS) solution is given by 
\begin{equation}\label{eqn:ols}
\thetahat= (XX\ic)^{-1}XY=\bigl(\sum\nolimits_{i\in [n]} x_ix_i\ic\bigr)^{-1}\bigl(\sum\nolimits_{i\in[n]} y_ix_i\bigr).
\end{equation}
While there are many private estimators for $\theta$, here we want to show how to use STR to develop a private estimator with basic statistical analysis. 

\subsection{Formulating STR}
To construct STR, we first seek a tractable statistic $\beta(\calX)$
that controls the local sensitivity of $\thetahat$. Consider neighboring
datasets $\calX\sim\calX'$ that differ by replacing $(x_i,y_i)$ with
$(x_i',y_i')$. For the simplicity of discussion, we assume that $\|x_i\|\le R_x$, $|y_i|\le R_y=R_\theta R_x$,
and $\|\thetahat(\calX)\|\le R_\theta$,
while later we can relax these requirements using projections. 
The bounds $R_x$ and $R_\theta$
may be specified from domain knowledge or obtained using a DP scale
estimator. Whenever the least-squares estimators are well defined,
the normal equations give
\begin{align}
\Delta_{\thetahat}(\calX)
&\le
\max_{\calX'\sim\calX}
\|(X'(X')\ic)^{-1}\|
\left(
\|x_i'(x_i')\ic-x_ix_i\ic\|\,\|\thetahat(\calX)\|
+
\|y_i'x_i'-y_ix_i\|
\right) \notag\\
&\le
4R_x^2R_\theta
\max_{\calX'\sim\calX}
\frac{1}{\lambda_p(X'(X')\ic)}.
\label{eqn:perturbOLS}
\end{align}
Although the exact local sensitivity is difficult to compute even for
this regression problem, \eqref{eqn:perturbOLS} suggests the regularity
statistic
\(
\beta(\calX)=\lambda_p(XX\ic).
\)

We next use $\beta$ to select a sensitivity threshold. By Weyl's
inequality, replacing one observation changes $\beta$ by at most
$2R_x^2$:
\begin{equation}
\label{eqn:weyl}
|\beta(\calX)-\beta(\calX')|
\le
\|x_ix_i\ic-x_i'(x_i')\ic\|
\le 2R_x^2.
\end{equation}
Fix $c_0>0$ and set
\(
\alpha=\frac{4R_x^2R_\theta}{c_0n}.
\)
If $\beta(\calX)\ge c_0n+2R_x^2$, then every neighboring dataset
satisfies $\lambda_p(X'(X')\ic)\ge c_0n$. Equation~\eqref{eqn:perturbOLS}
therefore implies $\Delta_{\thetahat}(\calX)\le\alpha$.

Finally, since $\beta$ is $2R_x^2$-Lipschitz, we define
\begin{equation}
\label{eqn:gammaOLS}
\gamma(\calX)
=
(2R_x^2)^{-1}{
\bigl(\lambda_p(XX\ic)-c_0n-2R_x^2\bigr)_+
}.
\end{equation}
Then $\gamma$ is $1$-Lipschitz and
\(
\{\gamma(\calX)>0\}
\subseteq
\{\Delta_{\thetahat}(\calX)\le\alpha\}.
\)
Hence, $\gamma$ is an $\alpha$-safety score. Its computation requires only the
smallest eigenvalue of $XX\ic$.

Under suitable conditions on the design, $\lambda_p(XX\ic)$ is of order
$n$ with high probability \citep{vershynin2010introduction}. Choosing
$c_0$ below the corresponding asymptotic lower bound makes
$\gamma(\calX)>0$ with high probability for sufficiently large $n$.
Thus, $\gamma(\calX)$ is typically positive while the sensitivity
guarantee used by STR holds for every dataset with positive safety score.

Based on $\alpha$ and $\alpha$-safety score $\gamma$, we finalize the STR estimator in Algorithm \ref{alg:ePTROLS}. 
In the algorithm, we enforce $\|x_i\|\leq R_x,  \|\thetahat\|\leq R_\theta, |y_i|\leq R_y=R_\theta R_x$ by projections. 

\begin{algorithm}
\caption{Linear regression with STR (STR-OLS)}\label{alg:ePTROLS}
\begin{algorithmic}[1]
\Require Data $\calX$, tolerance $\varepsilon,\delta$, prefixed bounds $R_x, R_\theta, c_0$.
\Ensure An $(\varepsilon,\delta)$-DP output $\thetatilde_\gamma(\calX)$
\State Let $x_i=\Pi_{R_x}(x_i),y_i=\Pi_{R_xR_\theta}(y_i)$.
\State Compute projected OLS $\thetahat=\Pi_{R_\theta}((XX\ic)^{-1}XY)$.
\State Apply STR on $\thetahat$ with $\gamma$ as in \eqref{eqn:gammaOLS} and $\alpha={4R_x^2R_\theta}/{nc_0}$ to have the release $\tilde{\theta}_{\gamma}(\calX)$.
\end{algorithmic}
\end{algorithm}

\begin{thm}
\label{thm:LRDPcheck}
The output of Algorithm~\ref{alg:ePTROLS} is $(\varepsilon,\delta)$-DP when $\varepsilon\in (0,2)$. 
\end{thm}

\subsection{Performance analysis}
We then discuss the performance of STR-OLS, including the error bound and consistency. Below are our model assumptions to facilitate such discussions. 
\begin{aspt}
\label{aspt:LR}
$p\geq 4$, 
$x_{[n]}$ are i.i.d. samples from $\mathcal{N}(0,\Sigma)$, and $y_i$ is generated from $y_i=\theta\ic x_i+\xi_i$ with $\xi_i$ being independent samples from $\mathcal{N}(0,\sigma_\xi^2)$.
In addition, assume the covariance matrix $\Sigma$ is well-conditioned, so that there is a constant $C_x > 0$, 
\[
C_x^{-2}\leq \lambda_p(\Sigma)\leq \cdots\leq \lambda_1(\Sigma)\leq C^2_x.
\]
Furthermore, there exists \(R_\theta>0\) that may or may not depend on $p$, such that
$\|\theta\|\le R_\theta$ and $\sigma_\xi\le \frac{1}{2}C_xR_\theta\sqrt p.$
\end{aspt}

We set $R_x =  2C_x\sqrt{p\log n}$ and $c_0\leq 1/{(4C_x^2)}$. Under Assumption \ref{aspt:LR}, the original data point $x_i$ is kept with high probability.
By random matrix theory, if the dimension $p<\frac{n}{64C_x^4\log n}$, then with high probability  
$\lambda_{p}(XX\ic)\geq \frac34 nC_x^{-2}$, and so
\[
\gamma(\calX)=\frac{1}{2 R_x^2}(\lambda_p(XX\ic)-c_0 n-2R_x^2)_+
\geq \frac{n}{32pC_x^4\log n}.
\]
Using the STR algorithm in Algorithm \ref{alg:ePTR}, Theorem \ref{thm:ePTRperformance} indicates that the non-release loss is negligible when $\varepsilon\gg \frac{p\log n}{n}\log(1/\delta)$. Rigorous derivations give Theorem \ref{thm:LRDP} below. 
\begin{thm}
\label{thm:LRDP}
Suppose Assumption \ref{aspt:LR} holds. 
For STR-OLS with $R_x=2C_x\sqrt{p\log n}$, $c_0\leq \frac{1}{4C_x^2}$, and $\bot=0$, with privacy budget 
$\delta\leq n^{-3}$ and $64pC_x^4(1+\frac{2}{\varepsilon}\log(\frac{1}{\delta})) \leq {n}/{\log n}$,  the expected mean square error of $\tilde{\theta}_{\gamma}$ is bounded by  
\[
L_\theta(\thetatilde_\gamma):=\E_{\theta}[\|\thetatilde_\gamma(\calX)-\theta\|^2]\leq C\biggl(\sigma_{\xi}^2 C_x^2\frac{p}{n}+R_\theta^2\frac{p^3 (\log n)^2\log (1/\delta)}{\varepsilon^2 n^2}\biggr).
\]
\end{thm}
Theorem \ref{thm:LRDP} shows that the error bound is at the order of $O(p/n)$ and $O(p^3/n^2\varepsilon^2)$ when both $C_x$ and $R_{\theta}$ are at constant order. 
\cite{cai2021cost} considers the case that $\lambda_i(\Sigma)\asymp 1/p$ for $i \in [p]$, and set up the minimax error bound that $\E[(\thetatilde-\theta)\ic\Sigma (\thetatilde-\theta)]\gtrsim\frac{p}{n}+\frac{p^2}{\varepsilon^2 n^2}$ for any $(\varepsilon,\delta)$-DP estimator $\thetatilde$. Furthermore, they suggest a noisy gradient descent method to achieve this rate, requiring that $n\gg p^{3/2}(\log n)^{3/2}/\varepsilon$.

Now we consider STR-OLS in this setting. Rescale $x^r_i = \sqrt{p} x_i$ and $\theta^r = \theta/\sqrt{p}$, then Assumption \ref{aspt:LR} holds with $R_{\theta} = O(1/\sqrt{p})$ and $C_x, \sigma_{\xi}$ are at constant order. 
Theorem \ref{thm:LRDP} shows that our STR-OLS estimator has a loss of $O(\frac{p}{n} + \frac{p^2(\log n)^3}{n^2\varepsilon^2})$ in this setting assuming $\delta\asymp n^{-3}$, which matches the optimal loss up to the $\log$ factors. But for STR-OLS, we only require the sample size $n \gg p(\log n)^2/\varepsilon$, instead of $n \gg p^{3/2}(\log n)^{3/2}/\varepsilon$.
Using the standard PTR, \cite{liu2022differential} has a similar rate to ours, but it is not computationally friendly.  

\subsection{Numerical simulation}\label{subsec:simu_linear}

We evaluate the proposed STR method under the mentioned linear regression model. Generate the covariates $x_i\sim {\calN}(0,I_p)$ and the response
$y_i=x_i^\top\theta+\xi_i$, where $\xi_i\sim \calN(0,1)$ and the underlying regression coefficient vector
$\theta=\frac{(1,1/2,\ldots,1/p)^\top}{\|(1,1/2,\ldots,1/p)^\top\|_2}$. Throughout, we set \(p=5\).

We compare five methods: the non-private estimator, the proposed STR-OLS estimator, the private Johnson--Lindenstrauss projection method (DPJL) by \cite{sheffet2017differentially}, the private gradient-descent method (DPGD) by \cite{cai2021cost}, and the functional mechanism (FM) as seen in \cite{zhang2012regression}. For each replication, all methods are trained on the same $n$ samples and evaluated on the same $m$ independent test samples. We report the mean squared error $\frac{1}{m}\sum_{i=1}^m (\hat{y}_i - y_i)^2$ on the test set. Throughout, $m=10{,}000$ and $\delta=0.01$. 

\begin{figure}[htbp]
    \centering
    \begin{subfigure}[t]{0.4\textwidth}
        \centering
        \includegraphics[width=\linewidth]{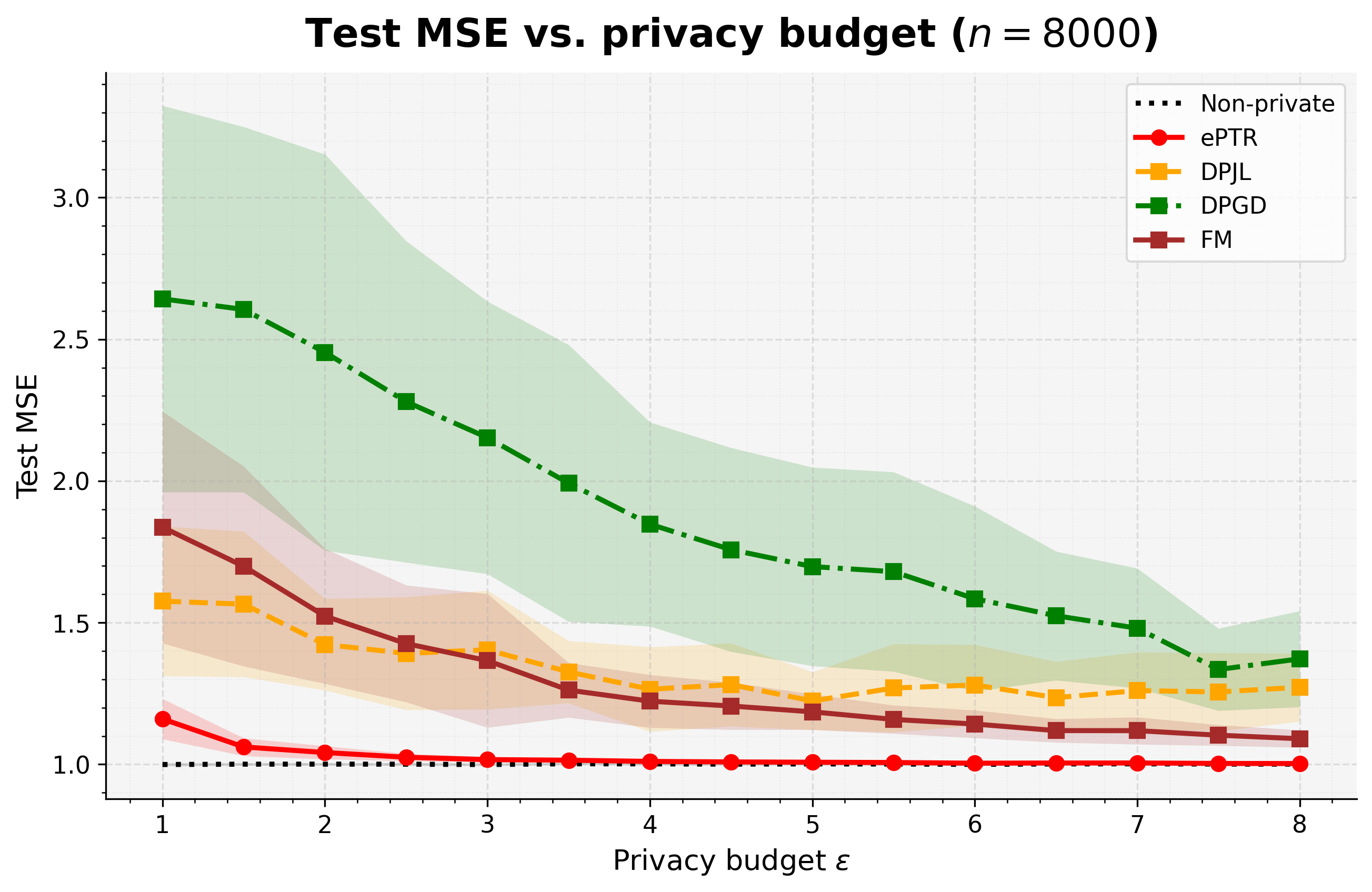}
    \end{subfigure}
    \begin{subfigure}[t]{0.4\textwidth}
        \centering
        \includegraphics[width=\linewidth]{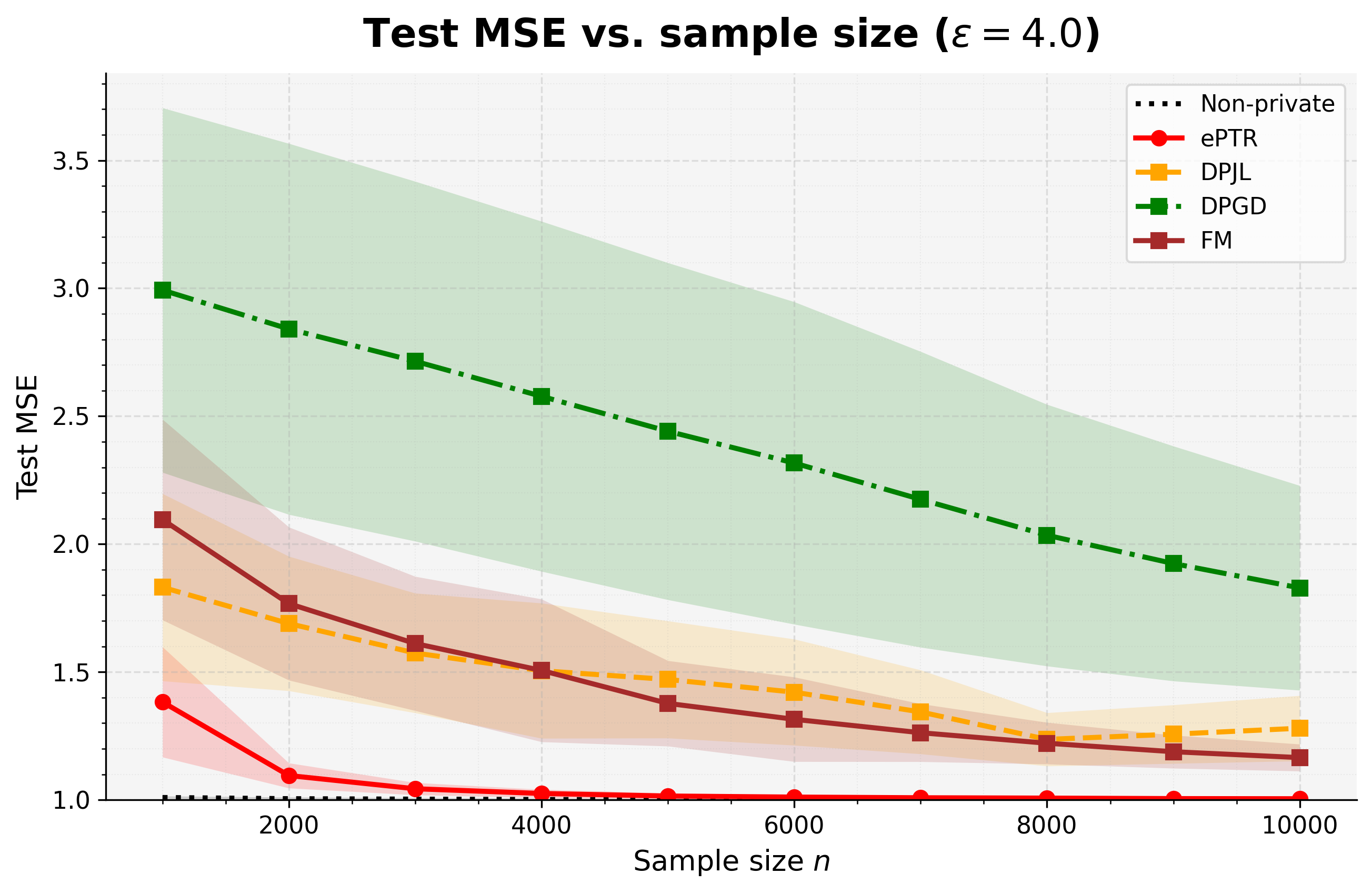}
    \end{subfigure}

    \caption{MSE on simulated data under varying privacy budget $\varepsilon$ and sample size $n$.}
    \label{fig:linear_all}
\end{figure}

We vary the privacy budget $\varepsilon \in \{1,1.5,2,\ldots,8\}$ with $n=8000$, 
and vary sample size $n\in\{1000,\ldots,10000\}$ in increments of $1{,}000$ with $\varepsilon=4.0$. For STR-OLS, we use $R_x=4$ and $R_\theta=1$ throughout. We set $c_0=0.4$ when varying $\varepsilon$ and $c_0=0.5$ when varying $n$. When $\varepsilon>2$, we verify the same Gaussian mechanism is DP using \cite{balle2018improving}.
Figure~\ref{fig:linear_all} summarizes the results across 500 replicates. Our STR-OLS approach consistently outperforms other private methods significantly. It rapidly converges to the non-private benchmark when the privacy budget $\varepsilon \geq 1.5$ or the sample size $n \geq 2{,}000$, while other private methods need a larger privacy budget or sample size. It shows the efficiency of our STR framework. 

\section{Kernelized regression}\label{sec:KR}
Nonparametric regression is a fundamental problem. Considering training data pairs $z_i=(x_i,y_i)$, suppose $y_i=f(x_i)+\xi_i$, where $x_i$ are i.i.d. samples from a density $\mu$ on $\mathbb{R}^d$ and $\xi_i\sim \mathcal{N}(0,\sigma_\xi^2)$. The goal is to estimate the function value $f(x_0)$ at a point of interest $x_0$. %

While there are plenty of works on nonparametric estimation, we focus on a simple and classical one: kernelized regression. 
The simplest estimator is the Nadaraya--Watson (NW) estimator \citep{wasserman2006all}, which is given by 
\[
\hat{f}(x_0)=\frac{1}{d_{\sigma}(x_0,\calX)}\sum_{i\in [n]} K_\sigma(x_0,x_i)y_i,\quad \text{where }d_{\sigma}(x_0,\calX):=\sum_{i\in [n]} K_\sigma(x_0,x_i).
\]
Here, $K_{\sigma}(\cdot, \cdot)$ is a kernel function with bandwidth $\sigma$ and $d_{\sigma}$ is often called the degree function. The choice of kernel $K$ depends on the smoothness of the function $f$ and density $\mu$. 
More discussions are left to the supplementary material. 

\subsection{Formulating STR}\label{subsec:ePTR_KR}
To construct STR for kernel regression, we first seek a tractable
statistic $\beta(\calX)$ that controls the local sensitivity of
$\fhat(x_0)$. Assume that
$|K_\sigma|\le \sigma^{-d}C_K$ for a constant $C_K>0$. For example,
the Gaussian kernel satisfies this condition with
$C_K=2$. We also assume that
$|\fhat(x_0)|,|y_i|\le R_f$ for a known $R_f$; the bound on the
responses can be enforced by projection preprocessing. For neighboring
datasets $\calX\sim\calX'$ that differ by replacing $(x_i,y_i)$ with
$(x_i',y_i')$, the perturbation satisfies
\begin{align}
&|\fhat(x_0,\calX)-\fhat(x_0,\calX')|\notag\\
&\quad \le
\frac{|K_\sigma(x_0,x_i)y_i
        -K_\sigma(x_0,x_i')y_i'|}
     {|d_\sigma(x_0,\calX')|} +
\frac{|K_\sigma(x_0,x_i)
        -K_\sigma(x_0,x_i')|}
     {|d_\sigma(x_0,\calX')|}
|\fhat(x_0,\calX)| \notag\\
&\quad \le
\frac{4C_KR_f}
     {\sigma^d |d_\sigma(x_0,\calX')|}.
\label{eqn:perturbKernel}
\end{align}
The denominator suggests the regularity
statistic
\(
\beta(\calX)=|d_\sigma(x_0,\calX)|.
\)

We next choose a sensitivity threshold using the stability of $\beta$.
For any neighboring $\calX'$,
\(|\beta(\calX')-\beta(\calX)|
\leq
|K_\sigma(x_0,x_i')-K_\sigma(x_0,x_i)|
\le 2\sigma^{-d}C_K.\)
Fix a $c_0>0$, which may depend on $\sigma$, and set
\(
\alpha=\frac{4C_KR_f}{\sigma^dc_0n}.
\)
If $\beta(\calX)\ge c_0n+2\sigma^{-d}C_K$, then
$|d_\sigma(x_0,\calX')|\ge c_0n$ for every neighboring $\calX'$.
Equation~\eqref{eqn:perturbKernel} therefore gives
$\Delta_{\fhat(x_0)}(\calX)\le\alpha$.

Finally, we can find that $\beta$ is
$2\sigma^{-d}C_K$-Lipschitz, and therefore we consider
\begin{equation}
\label{eqn:gammaKR}
\gamma(\calX)
=
(2\sigma^{-d}C_K)^{-1}{
\bigl(|d_\sigma(x_0,\calX)|-c_0n-2\sigma^{-d}C_K\bigr)_+
}.
\end{equation}
Then $\gamma$ is $1$-Lipschitz and
\(
\{\gamma(\calX)>0\}
\subseteq
\{\Delta_{\fhat(x_0)}(\calX)\le\alpha\},
\)
so $\gamma$ is an $\alpha$-safety score.

Under conditions ensuring that
$d_\sigma(x_0,\calX)/n$ is bounded away from zero with high
probability, $c_0$ can be chosen so that $\gamma(\calX)>0$ with high
probability. The degree and $\gamma(\calX)$ are directly
computable, yielding the STR kernel regression estimator in
Algorithm~\ref{alg:ePTRKR}.

\begin{algorithm}
\caption{Nadaraya--Watson with STR (STR-NW)}\label{alg:ePTRKR}
\begin{algorithmic}[1]
\Require Data $\calX$, tolerance $\varepsilon,\delta$, prefixed bounds $R_f, c_0$ and kernel upper bound $C_K$. 
\Ensure An $(\varepsilon,\delta)$-DP output $\fhat_\gamma(x_0)$
\State Let $y_i=\Pi_{R_f}(y_i)$.
\State Compute projected kernel regression 
\[
\fhat(x_0)=\Pi_{R_f}\bigl( \sum\nolimits_{i\in [n]}K_\sigma (x_0,x_i)y_i/d_{\sigma}(x_0,\calX)\bigr).
\]
\State Apply STR on $\fhat(x_0)$ with $\alpha=\frac{4 R_f C_K}{\sigma^d c_0 n}$ and safety score $\gamma(\calX)$ in \eqref{eqn:gammaKR}.  
\State Let the output be $\fhat_\gamma(x_0)$. Return it as the private estimate. 
\end{algorithmic}
\end{algorithm}

\begin{theorem}
\label{thm:KRDP}
Suppose $|K_\sigma|\leq \sigma^{-d}C_K$. Then Algorithm~\ref{alg:ePTRKR} is $(\varepsilon,\delta)$-DP when $\varepsilon \in (0,2)$.
\end{theorem}

Due to space limit, we briefly discuss the main consistency results here and leave the technical details to Appendix A of the supplementary material. 
Suppose that $K_{\sigma}$ is a regular radial basis kernel with bandwidth $\sigma$ and order $s$, meaning that its rescaled kernel has vanishing moments up to degree $s-1$ and hence induces a smoothing bias of order $\sigma^s$ for $s$-smooth functions.
Under some regular conditions, choosing the bandwidth
$\sigma \asymp \max\Big\{n^{-\frac{1}{2s+d}},\ \Big(\tfrac{n\varepsilon}{\sqrt{\log(1/\delta)}}\Big)^{-\frac{1}{s+d}}\Big\}$,
and provided $n\varepsilon \gtrsim (\log(1/\delta))^{1+d/(2s)}$ and \(n\varepsilon/\sqrt L\gg(\log n)^{(s+d)/2}\) with $L = \log(1/\delta)$, Algorithm~\ref{alg:ePTRKR}
achieves the mean square error
$\widetilde{O}\big(n^{-\frac{2s}{2s+d}}+(n\varepsilon)^{-\frac{2s}{s+d}}\big)$;
details in Corollary A.5. The first term is the classical nonparametric rate and the
second is the cost of privacy. When $\varepsilon \lesssim n^{-s/(2s+d)}$, the bandwidth is
enlarged beyond the non-private choice. This rate matches the lower bound
of \citet{cai2024optimalb} up to polylogarithmic factors when $d=1$.

\subsection{California housing kernel regression}
We evaluate the proposed STR procedures on the California Housing dataset,
originally derived from the 1990 California census \citep{pace1997sparse}. 
The dataset contains $20{,}640$ California
census-block records. 
We consider kernel regression with median house value as the response, measured in units of $100{,}000$ dollars. The selected predictors summarize local income level, geographic location, and housing-stock age, which are important factors for local housing-value prediction. Although the data are aggregated at the census-block level, it is natural that the participants would like their income to be protected. Specifically, we use median income, latitude, longitude, and housing age, so $d=4$. These predictors are clipped
to fixed public ranges and linearly rescaled to $[0,1]$. The response is clipped to $[0,5]$ and transformed as $(Y-2.5)/2.5$, giving $R_f=1$ on the transformed scale. We use a Gaussian kernel with bandwidth $\sigma=0.5n^{-1/(d+4)}$ and set $c_0=0.3$.

\begin{figure}[htbp]
    \centering
    \includegraphics[width=0.82\textwidth, height = 1.5in]{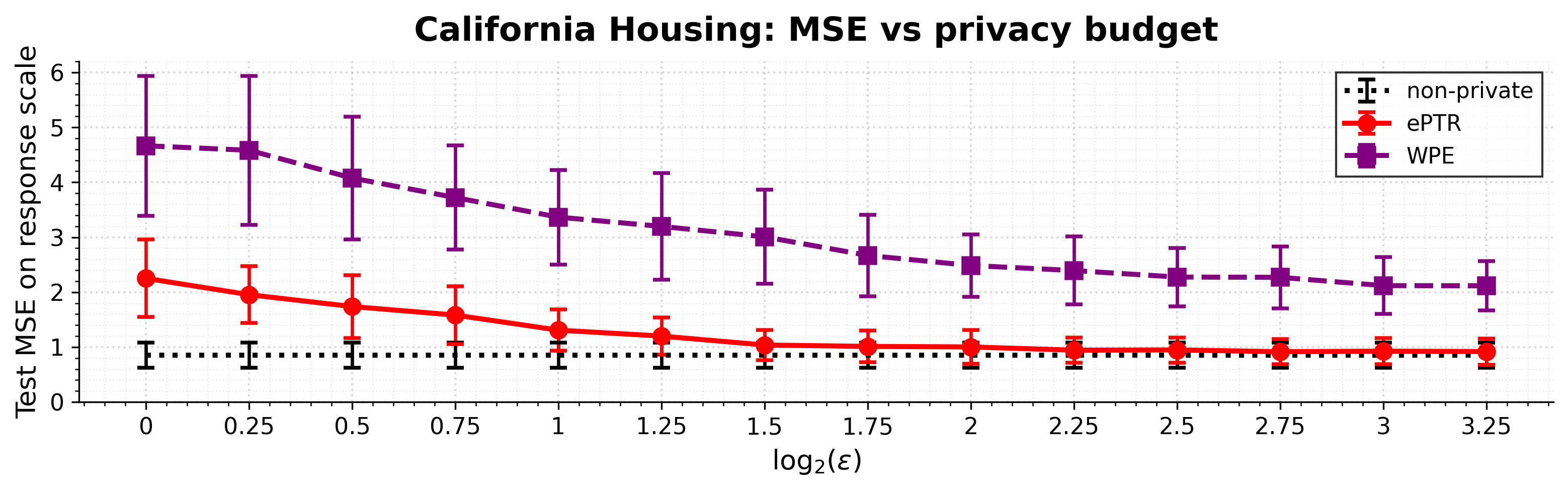}
    \caption{California Housing data: test MSE versus $\log_2(\varepsilon)$
    for kernel regression.}
    \label{fig:real_california}
\end{figure}

We consider the non-private NW estimator, our STR-NW estimator defined in  Algorithm \ref{alg:ePTRKR}, and the wavelet-based private estimator (WPE) in \cite{cai2024optimalb}. 
For each replication, we split the data into a $20\%$ training sample and a held-out $80\%$ testing sample. All methods are fitted only on the training sample, and their performance is evaluated on the same held-out testing sample. We consider the MSE of the point estimate at a random testing sample. Numerically this is equivalent to finding the MSE averaged over all testing samples. 50 replications are implemented to further average out the DP noise effect. The results are shown in Figure~\ref{fig:real_california}.
 Our STR-NW algorithm consistently outperforms WPE. Furthermore, it converges to the non-private benchmark when the privacy budget $\varepsilon \geq 1.5$, which corresponds to a moderate privacy regime. When $\varepsilon>2$, we verify the same Gaussian mechanism is DP using \cite{balle2018improving}.

We also evaluate our STR-NW estimator on simulated datasets, where our method has shown overall improved performance. The details can be found in the supplementary materials. 

\section*{Acknowledgements}
The authors used ChatGPT (OpenAI) to assist with language
refinement. The authors reviewed the resulting text and take full responsibility for the article.

\section*{Data Availability}

The code, datasets, and reproduction guidance are available in the repository
\url{https://github.com/taoshen2022/Score-Test-Release--}.

\clearpage
\begingroup
\small
\setlength{\bibsep}{3pt}
\bibliographystyle{abbrvnat}
\bibliography{ref}

@inproceedings{guhathakurta2013differentially,
  title     = {Differentially Private Feature Selection via Stability Arguments, and the Robustness of the Lasso},
  author    = {{Guha Thakurta}, Abhradeep and Smith, Adam},
  booktitle = {Proceedings of the 26th Annual Conference on Learning Theory},
  series    = {Proceedings of Machine Learning Research},
  volume    = {30},
  pages     = {819--850},
  editor    = {Shalev-Shwartz, Shai and Steinwart, Ingo},
  publisher = {PMLR},
  year      = {2013},
  url       = {https://proceedings.mlr.press/v30/Guha13.html}
}

@article{dwork2014algorithmic,
  title={The algorithmic foundations of differential privacy},
  author={Dwork, Cynthia and Roth, Aaron and others},
  journal={Foundations and trends{\textregistered} in theoretical computer science},
  volume={9},
  number={3--4},
  pages={211--407},
  year={2014},
  publisher={Now Publishers, Inc.}
}

@article{bu2020deep,
  title={Deep learning with {Gaussian} differential privacy},
  author={Bu, Zhiqi and Dong, Jinshuo and Long, Qi and Su, Weijie J},
  journal={Harvard data science review},
  volume={2020},
  number={23},
  pages={10--1162},
  year={2020}
}

@article{alabi2020differentially,
  title={Differentially private simple linear regression},
  author={Alabi, Daniel and McMillan, Audra and Sarathy, Jayshree and Smith, Adam and Vadhan, Salil},
  journal={arXiv preprint arXiv:2007.05157},
  year={2020}
}

@inproceedings{erlingsson2014rappor,
  title={Rappor: Randomized aggregatable privacy-preserving ordinal response},
  author={Erlingsson, {\'U}lfar and Pihur, Vasyl and Korolova, Aleksandra},
  booktitle={Proceedings of the 2014 ACM SIGSAC conference on computer and communications security},
  pages={1054--1067},
  year={2014}
}

@book{wasserman2006all,
  title={All of nonparametric statistics},
  author={Wasserman, Larry},
  year={2006},
  publisher={Springer}
}

@article{donoho1998minimax,
  title={Minimax estimation via wavelet shrinkage},
  author={Donoho, David L and Johnstone, Iain M},
  journal={The annals of Statistics},
  volume={26},
  number={3},
  pages={879--921},
  year={1998},
  publisher={Institute of Mathematical Statistics}
}

@article{zhang2012regression,
  title={Functional Mechanism: Regression Analysis under Differential Privacy},
  author={Zhang, Jun and Zhang, Zhenjie and Xiao, Xiaokui and Yang, Yin and Winslett, Marianne},
  journal={Proceedings of the VLDB Endowment},
  volume={5},
  number={11},
  year={2012}
}

@inproceedings{sheffet2017differentially,
  title={Differentially private ordinary least squares},
  author={Sheffet, Or},
  booktitle={International Conference on Machine Learning},
  pages={3105--3114},
  year={2017},
  organization={PMLR}
}

@article{yeh2009comparisons,
  title={The comparisons of data mining techniques for the predictive accuracy of probability of default of credit card clients},
  author={Yeh, I-Cheng and Lien, Che-hui},
  journal={Expert systems with applications},
  volume={36},
  number={2},
  pages={2473--2480},
  year={2009},
  publisher={Elsevier}
}

@article{cortez2009modeling,
  title={Modeling wine preferences by data mining from physicochemical properties},
  author={Cortez, Paulo and Cerdeira, Antonio and Almeida, Fernando and Matos, Telmo and Reis, Jose},
  journal={Decision Support Systems},
  volume={47},
  number={4},
  pages={547--553},
  year={2009},
  publisher={Elsevier}
}

@article{cai2024optimalb,
  title={Optimal federated learning for nonparametric regression with heterogeneous distributed differential privacy constraints},
  author={Cai, T Tony and Chakraborty, Abhinav and Vuursteen, Lasse},
  journal={arXiv preprint arXiv:2406.06755},
  year={2024}
}

@article{wang2022renyi,
  title={R{\'e}nyi differential privacy of propose-test-release and applications to private and robust machine learning},
  author={Wang, Jiachen T and Mahloujifar, Saeed and Wang, Shouda and Jia, Ruoxi and Mittal, Prateek},
  journal={Advances in Neural Information Processing Systems},
  volume={35},
  pages={38719--38732},
  year={2022}
}

@inproceedings{nissim2007smooth,
  title={Smooth sensitivity and sampling in private data analysis},
  author={Nissim, Kobbi and Raskhodnikova, Sofya and Smith, Adam},
  booktitle={Proceedings of the thirty-ninth annual ACM symposium on Theory of computing},
  pages={75--84},
  year={2007}
}

@article{bun2019average,
  title={Average-case averages: Private algorithms for smooth sensitivity and mean estimation},
  author={Bun, Mark and Steinke, Thomas},
  journal={Advances in Neural Information Processing Systems},
  volume={32},
  year={2019}
}

@inproceedings{gonem2018smooth,
  title={Smooth sensitivity based approach for differentially private PCA},
  author={Gonem, Alon and Gilad-Bachrach, Ram},
  booktitle={Algorithmic Learning Theory},
  pages={438--450},
  year={2018},
  organization={PMLR}
}

@article{auddy2025minimax,
  title={Minimax and adaptive transfer learning for nonparametric classification under distributed differential privacy constraints},
  author={Auddy, Arnab and Cai, T Tony and Chakraborty, Abhinav},
  journal={Journal of the Royal Statistical Society Series B: Statistical Methodology},
  pages={qkaf070},
  year={2025},
  publisher={Oxford University Press UK}
}

@inproceedings{abadi2016deep,
  title={Deep learning with differential privacy},
  author={Abadi, Martin and Chu, Andy and Goodfellow, Ian and McMahan, H Brendan and Mironov, Ilya and Talwar, Kunal and Zhang, Li},
  booktitle={Proceedings of the 2016 ACM SIGSAC conference on computer and communications security},
  pages={308--318},
  year={2016}
}

@inproceedings{mcmahan2017learning,
  title={Learning differentially private recurrent language models},
  author={McMahan, H Brendan and Ramage, Daniel and Talwar, Kunal and Zhang, Li},
  booktitle={Proceedings of the 2018 International Conference on Learning Representations},
  year={2018}
}

@inproceedings{vaidya2013differentially,
  title={Differentially private {Naive} {Bayes} classification},
  author={Vaidya, Jaideep and Shafiq, Basit and Basu, Anirban and Hong, Yuan},
  booktitle={2013 IEEE/WIC/ACM International Joint Conferences on Web Intelligence (WI) and Intelligent Agent Technologies (IAT)},
  volume={1},
  pages={571--576},
  year={2013},
  organization={IEEE}
}

@incollection{tsybakov2008nonparametric,
  title={Nonparametric estimators},
  author={Tsybakov, Alexandre B},
  booktitle={Introduction to Nonparametric Estimation},
  pages={1--76},
  year={2008},
  publisher={Springer}
}

@article{zafarani2020differentially,
  title={Differentially private naive bayes classifier using smooth sensitivity},
  author={Zafarani, Farzad and Clifton, Chris},
  journal={arXiv preprint arXiv:2003.13955},
  year={2020}
}

@article{vershynin2010introduction,
  title={Introduction to the non-asymptotic analysis of random matrices},
  author={Vershynin, Roman},
  journal={arXiv preprint arXiv:1011.3027},
  year={2010}
}

@inproceedings{balle2018improving,
  title     = {Improving the {G}aussian Mechanism for Differential Privacy: Analytical Calibration and Optimal Denoising},
  author    = {Balle, Borja and Wang, Yu-Xiang},
  booktitle = {Proceedings of the 35th International Conference on Machine Learning},
  series    = {Proceedings of Machine Learning Research},
  volume    = {80},
  pages     = {394--403},
  publisher = {PMLR},
  year      = {2018}
}

@article{cai2021cost,
  title={The cost of privacy: Optimal rates of convergence for parameter estimation with differential privacy},
  author={Cai, T Tony and Wang, Yichen and Zhang, Linjun},
  journal={The Annals of Statistics},
  volume={49},
  number={5},
  pages={2825--2850},
  year={2021},
  publisher={Institute of Mathematical Statistics}
}

@article{dong2022gaussian,
  title={Gaussian differential privacy},
  author={Dong, Jinshuo and Roth, Aaron and Su, Weijie J},
  journal={Journal of the Royal Statistical Society Series B: Statistical Methodology},
  volume={84},
  number={1},
  pages={3--37},
  year={2022},
  publisher={Oxford University Press}
}

@inproceedings{abowd2018us,
  title={The {US} {Census} {Bureau} adopts differential privacy},
  author={Abowd, John M},
  booktitle={Proceedings of the 24th ACM SIGKDD international conference on knowledge discovery \& data mining},
  pages={2867--2867},
  year={2018}
}

@inproceedings{dwork2006differential,
  title={Differential privacy},
  author={Dwork, Cynthia},
  booktitle={International colloquium on automata, languages, and programming},
  pages={1--12},
  year={2006},
  organization={Springer}
}

@inproceedings{dwork2009differential,
  title={Differential privacy and robust statistics},
  author={Dwork, Cynthia and Lei, Jing},
  booktitle={Proceedings of the forty-first annual ACM symposium on Theory of computing},
  pages={371--380},
  year={2009}
}

@inproceedings{liu2022differential,
  title={Differential privacy and robust statistics in high dimensions},
  author={Liu, Xiyang and Kong, Weihao and Oh, Sewoong},
  booktitle={Conference on Learning Theory},
  pages={1167--1246},
  year={2022},
  organization={PMLR}
}

@article{brunel2020propose,
  title={Propose, test, release: Differentially private estimation with high probability},
  author={Brunel, Victor-Emmanuel and Avella-Medina, Marco},
  journal={arXiv preprint arXiv:2002.08774},
  year={2020}
}

@inproceedings{redberg2023generalized,
  title={Generalized ptr: User-friendly recipes for data-adaptive algorithms with differential privacy},
  author={Redberg, Rachel and Zhu, Yuqing and Wang, Yu-Xiang},
  booktitle={International Conference on Artificial Intelligence and Statistics},
  pages={3977--4005},
  year={2023},
  organization={PMLR}
}

@article{pace1997sparse,
  title={Sparse Spatial Autoregressions},
  author={Pace, R. Kelley and Barry, Ronald},
  journal={Statistics \& Probability Letters},
  volume={33},
  number={3},
  pages={291--297},
  year={1997},
  publisher={Elsevier}
}

@book{bayes,
  title     = {A Probabilistic Theory of Pattern Recognition},
  author    = {Devroye, Luc and Gy{\"o}rfi, L{\'a}szl{\'o} and Lugosi, G{\'a}bor},
  publisher = {Springer},
  year      = {1996}
}

@article{wang2026net,
  title={{NetPTR}: Optimal Differentially Private Spectral Community Detection on Sparse Networks},
  author={Wang, Wanjie and Shen, Tao},
  journal={arXiv preprint arXiv:2606.26145},
  year={2026}
}
\endgroup

\clearpage
\section*{Supplementary Material}
\addcontentsline{toc}{section}{Supplementary Material}
This supplementary material provides additional details for the main manuscript, including background and simulation results about kernelized regression, and the proofs for the proposed STR procedures. The numbering and labels follow the main manuscript unless stated otherwise.

\appendix
\numberwithin{equation}{section}
\numberwithin{figure}{section}
\numberwithin{table}{section}
\setcounter{equation}{0}
\setcounter{figure}{0}
\setcounter{table}{0}
\renewcommand{\ic}{^\intercal}
\section{Kernelized regression}
\subsection{A brief introduction to kernel estimation}
Most content of this part is not new and can be found in nonparametric estimation textbooks like \cite{tsybakov2008nonparametric}. We provide a compact version of them here for self-completeness.

We consider the Nadaraya–Watson (NW) estimator \citep{wasserman2006all}, which is given by 
\[
\hat{f}(x_0)=\frac{1}{d_{\sigma}(x_0,\calX)}\sum_{i\in [n]} K_\sigma(x_0,x_i)y_i,\quad \text{where }d_{\sigma}(x_0,\calX):=\sum_{i\in [n]} K_\sigma(x_0,x_i),
\]
The choice of the kernel depends on the smoothness of the function and $\mu$. 
For example, when $f$ and $\mu$ are up to second order smooth, it is a common choice to use a Gaussian kernel.
\begin{equation}
\label{eqn:Gaussiankernel}
    K_{\sigma}(x,x')=\frac{1}{\sqrt{2\pi\sigma^2}^d}\exp(-\frac{\|x-x'\|^2}{2\sigma^2}).
\end{equation}
The standard deviation $\sigma$ is often known as the bandwidth. 

To understand why the NW estimator works, it suffices to know that if $\mu,f$ are $\mathcal{C}^2$ functions,
then with high probability
\[
\frac{1}{n}\sum_{i\in [n]} K_\sigma(x_0,x_i)f(x_i)\approx f(x_0)\mu(x_0),
\]
and
\[
\frac{1}{n}d_{\sigma}(x_0,\calX)=\frac{1}{n}\sum_{i\in [n]} K_\sigma(x_0,x_i)\approx \mu(x_0),
\]
therefore their ratio will approximately be $f(x_0)$, the quantity of interest. What is left in the NW estimator $\fhat(x_0)$ are noises coming from $\xi_i$. But they can be managed by standard concentration inequalities.  

When $\mu$ and $f$ are of higher order $\mathcal{C}^s$ smoothness, the approximation above can be made more accurate. But this would require one to use some other kernels, of which the introduction is slightly  technical. For readers who are interested in $\mathcal{C}^2$ functions, this part can be skipped since the Gaussian kernel \eqref{eqn:Gaussiankernel} suffices. 

\begin{suppdefn}
We say $f\in \mathcal{C}^s$ at $x$, if $\partial^{\kappa}f(x)$ exists for any multi-index $\kappa$ such that $|\kappa|_1\leq s$. Moreover for some $r_0$ and $L_f$, the multivariate Taylor expansion holds for all $y: \|y-x\|\leq r_0$
\[
\biggl|f(y)-f(x)-\sum_{j=1}^{s-1} \frac{1}{j!} \nabla^j f(x)[y-x,\ldots,y-x]\biggr|\leq L_f \|y-x\|^s. 
\]
\end{suppdefn}
Intuitively, a smoother function will be easier to estimate. To leverage the order $s$ smoothness, one needs to use an order $s$ kernel, which is defined as below (Note that in \citep{tsybakov2008nonparametric}, it is defined as order $s-1$ instead).
\begin{suppdefn}
$K_\sigma(x,x')$ is a regular radial basis kernel with bandwidth $\sigma$ and order $s$ if $K_{\sigma}(x,x')=\frac{1}{\sigma^d}K(\frac{\|x-x'\|}{\sigma})$ for some symmetric function $K$ such that 
\begin{equation}
\label{eqn:Ksorder}
\int_{\reals^d} K(\|u\|)\,du=1,\quad \int_{\reals^d} K(\|u\|)\,u^{\otimes j}\, du=0,\quad j=1,\ldots,s-1.   
\end{equation}
And there is a constant $C_K\geq 2$ so that
\[
|K(u)|\leq C_K\exp(-|u|^2/C_K).  
\]
\end{suppdefn}

It is easy to verify that the Gaussian kernel \eqref{eqn:Gaussiankernel} is a regular radial basis kernel with bandwidth $\sigma$ of order $2$. Another popular order $2$ kernel uses $K(v)=\frac{1}{V_d}1_{|v|\leq 1}$ where $V_d$ is the volume of the radius $1$ ball. Then the corresponding NW estimator $\fhat(x_0)$ will be averaging $y_i$ within the radius $\sigma$ ball around $x_0$.

For order $s\geq 3$ kernels, $K$ needs to take negative values but they exist for any fixed $s$.
Proposition 1.3 in \citep{tsybakov2008nonparametric} shows how to build $K$ when $d=1$ using Legendre polynomials. 
In general, $K$ can be constructed as a linear combination of basic functions such as polynomials or wavelets, with the coefficients chosen to satisfy \eqref{eqn:Ksorder}. Lemma~\ref{lem:buildK} gives such a construction using Gaussian densities.

When $f$ and $\mu$ are $\mathcal{C}^s$ at $x_0$, one classical result of kernel approximation is the following (see Proposition~\ref{pro:NW}):
\[
\biggl|\frac1n\sum_{i\in [n]} K_\sigma(x_0,x_i)f(x_i)- f(x_0)\mu(x_0)\biggr|=O_p\left(\sigma^s+\frac{(\log n)^{1+d/2}}{\sqrt{n\sigma^d}}\right).
\]
One special case is $f\equiv 1$, then with high probability
\begin{equation}
\label{eqn:dsigmatypical}
|d_\sigma(x_0,\calX)-n \mu(x_0)|=O\left(n\sigma^s+(\log n)^{1+d/2}\sqrt{n/\sigma^{d}}\right)=o(n)
\end{equation}
whenever $\sigma\to 0$ and $n\sigma^d/(\log n)^{2+d}\to\infty$.
This will be useful when we estimate the scale of $\gamma(\calX)$. 

\subsection{Performance analysis}
In order to evaluate the performance of STR-NW, we need to make assumptions on the data-generating process. We have discussed most of them during our derivation. We gather them below formally:
\begin{suppaspt}
\label{aspt:DGPKR}
For some constants $C_f,C_\mu,L_f,L_{\mu}, r_0$, the following hold
\begin{itemize}
    \item $x_i$ are i.i.d. samples from a distribution with density function $\mu(x)$ (w.r.t. Lebesgue measure).
 \item $y_i=f(x_i)+\xi_i$ where $\xi_i$ are i.i.d. samples from $\mathcal{N}(0,\sigma_{\xi}^2)$
 \item $|f(x)|\leq C_f$, $\mu(x)\leq C_\mu$, $\sigma_{\xi}\leq C_f$. 
 \item $\mu(x)$ and $f(x)$ are $\mathcal{C}^s$ at $x_0$ with regularity $L_f$ and $L_{\mu}$. 
\end{itemize}
\end{suppaspt}

Using Theorem \ref{thm:ePTRperformance}, we can predict the performance of STR-NW by checking the typical scale of $\gamma(\calX)$. 
Using \eqref{eqn:dsigmatypical}, we can find $\gamma(\calX)$ is of scale $\Omega(n\sigma^{d})$ if $\mu(x_0)>2c_0, n\sigma^{d}>1$.  This suggests that $\varepsilon\gg \frac{\log(1/\delta)}{n\sigma^d}$ will guarantee that the non-release loss is negligible. More formally, we have
\begin{supptheorem}
\label{thm:KR}
Under Assumption \ref{aspt:DGPKR}, suppose 
$K_\sigma(x,x')$ is a regular radial basis kernel with bandwidth $\sigma$ and order $s$, where $\sigma\to0$, $n\sigma^d/(\log n)^{2+d}\to\infty$, and $\sigma\sqrt{\log n}\to0$, with a fixed constant $c_0$ so that
$\mu(x_0)>2c_0+\frac{4}{n\sigma^d}C_K, \delta\leq \min\{n^{-3},\varepsilon\}$, $\bot=0$ and $R_f=C_f(1+\sqrt{8\log n})$, then when $n\sigma^d\mu(x_0)\geq 16C_K\big(1+\frac{2}{\varepsilon}\log(1/\delta)\big)$ and $n$ is sufficiently large, the following estimate holds
\[
\E\big[|\fhat_\gamma(x_0)-f(x_0)|^2\big]=O\left(\sigma^{2s}+\frac{(\log n)^{2+d}}{n \sigma^d}+\frac{\log n\,\log(1/\delta)}{\varepsilon^2(n\sigma^d)^2}\right).
\]
\end{supptheorem}
For kernelized estimation, the well known optimal scaling of bandwidth vs sample size is $\sigma=n^{-\frac{1}{d+2s}}$, which can be seen as the optimal tradeoff point between the bias and variance.  In the DP regime, we also need to balance the DP loss term. This leads to the following:
\begin{suppcor}
\label{cor:KR}
Let $L=\log(1/\delta)$ and $\sigma=\max\{n^{-\frac{1}{2s+d}},\,(n\varepsilon/\sqrt{L})^{-\frac{1}{s+d}}\}$. Under the same
assumptions as Theorem~\ref{thm:KR}, if in addition
$(n\varepsilon)^{\frac{s}{s+d}}L^{\frac{d}{2(s+d)}}\mu(x_0)\geq 16C_K(\varepsilon+2L)$ and $n\varepsilon/\sqrt{L}\gg(\log n)^{(s+d)/2}$, then
\[
\E\big[|\fhat_\gamma(x_0)-f(x_0)|^2\big]
=\widetilde{O}\Big(n^{-\frac{2s}{2s+d}}+(n\varepsilon)^{-\frac{2s}{s+d}}\Big).
\]
\end{suppcor}
\begin{proof}
Since $\sigma\ge n^{-1/(2s+d)}$, the variance term satisfies $\frac{1}{n\sigma^d}\le n^{-\frac{2s}{2s+d}}$;
since $\sigma\ge (n\varepsilon/\sqrt L)^{-1/(s+d)}$, we have $n\varepsilon\sigma^d\ge (n\varepsilon)^{\frac{s}{s+d}}L^{\frac{d}{2(s+d)}}$, so the privacy term satisfies
$\frac{L}{\varepsilon^2(n\sigma^d)^2}\le (n\varepsilon/\sqrt L)^{-\frac{2s}{s+d}}$; and the bias satisfies
$\sigma^{2s}=\max\{n^{-\frac{2s}{2s+d}},(n\varepsilon/\sqrt L)^{-\frac{2s}{s+d}}\}$. 
Finally, the additional condition gives $n\sigma^d\mu(x_0)\geq 16C_K(1+2L/\varepsilon)$, so the condition of
Theorem~\ref{thm:KR} holds. The claim follows.
\end{proof}
The first term is the classical nonparametric rate for $\mathcal{C}^s$ functions
\citep{donoho1998minimax}, and the second term is the cost of privacy; it dominates
when $\varepsilon\lesssim n^{-s/(2s+d)}$. When $d=1$, this matches the lower bound in
\citep{cai2024optimalb} up to logarithmic factors.
\subsection{Numerical simulation}

We evaluate the proposed STR method under the given nonparametric regression model and we take
\[
f(x)=\sin(2\pi x)+0.5\cos(4\pi x), \qquad \xi_i \sim N(0,\sigma_\xi^2),
\]
with \(\sigma_\xi=0.2\) and target point \(x_0=0.5\) for which \(f(x_0)=0.5\). 

We also consider three simulation regimes: varying the privacy budget $\varepsilon$, varying the training sample size $n$, and varying the concentration parameter $a$ to study the influence of local density. For the first two regimes, \(x_i\sim \mathrm{Unif}[0,1]\), and for the last one, we use \(x_i\sim \mathrm{Beta}(a,a)\) to vary the local concentration of observations around \(x_0\).
Performance is evaluated by the pointwise squared error \([\hat f(x_0)-f(x_0)]^2,\)
averaged over 500 Monte Carlo replications. We compare three methods: the non-private Nadaraya--Watson estimator, a wavelet-based private estimator (WPE) for nonparametric regression as seen in \citep{cai2024optimalb}, and the proposed STR estimator for which we use the Nadaraya--Watson estimator with Gaussian kernel and the bandwidth is set as \(0.2\,n^{-1/5}.\)

\begin{figure}[htbp]
    \centering
    \begin{subfigure}[t]{0.32\textwidth}
        \centering
        \includegraphics[width=\linewidth]{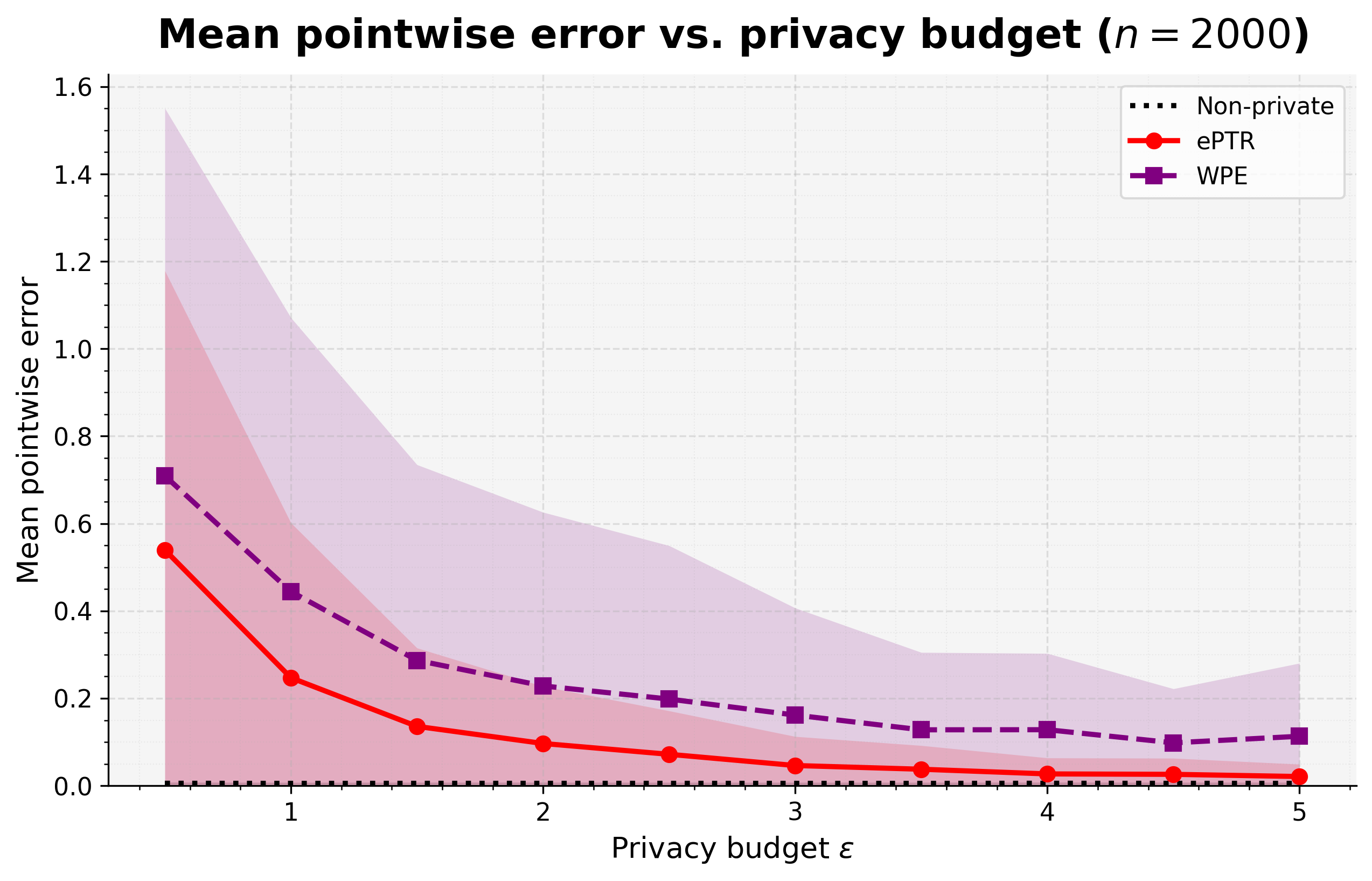}
    \end{subfigure}
    \begin{subfigure}[t]{0.32\textwidth}
        \centering
        \includegraphics[width=\linewidth]{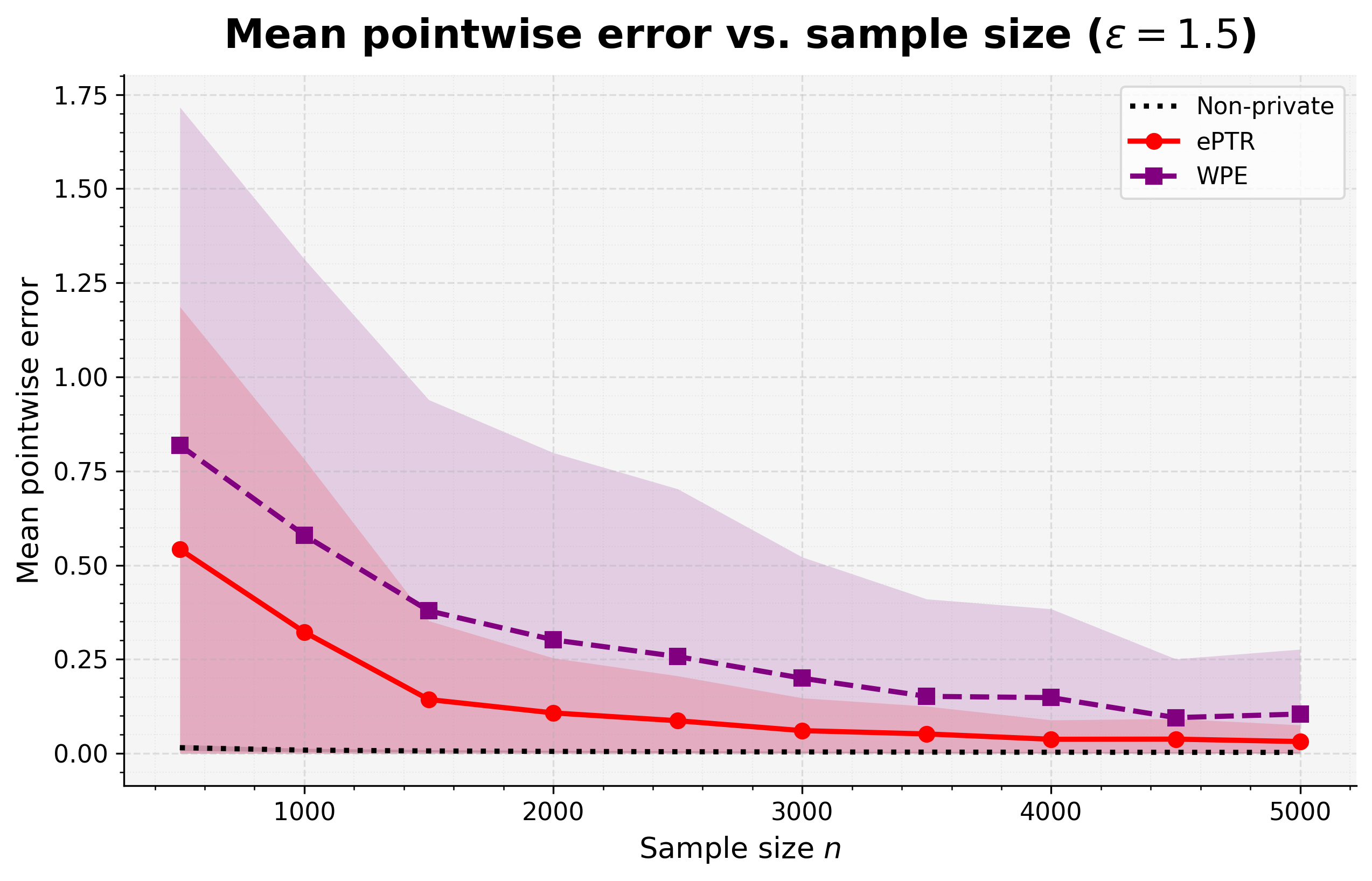}
    \end{subfigure}
    \begin{subfigure}[t]{0.32\textwidth}
        \centering
        \includegraphics[width=\linewidth]{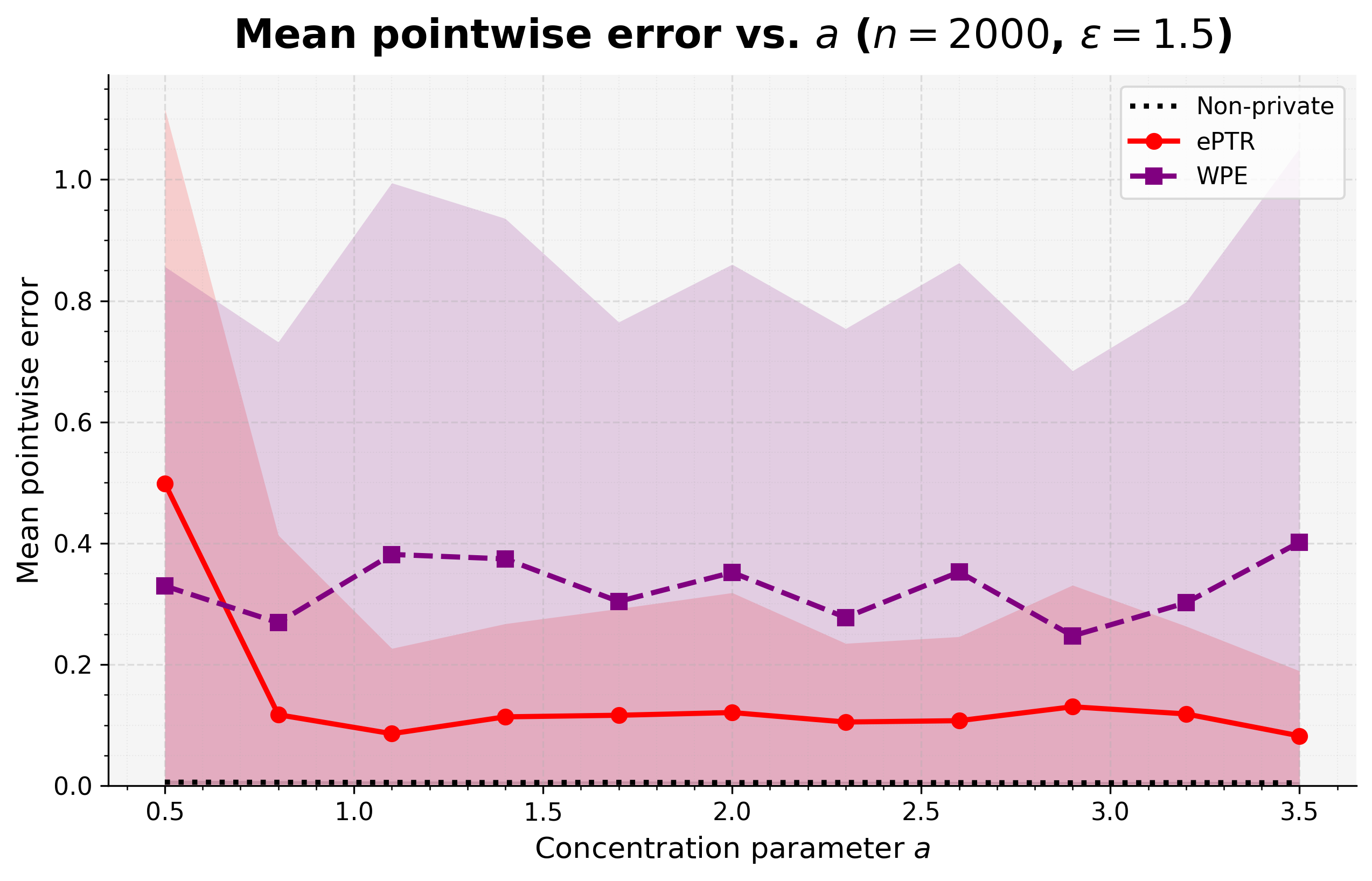}
    \end{subfigure}
    \caption{Simulation results for kernelized regression under varying privacy budget, sample size, and design local density. }
    \label{fig:kernel_sim_all}
\end{figure}

Figure~\ref{fig:kernel_sim_all} summarizes the simulation results. As expected, the non-private estimator achieves the smallest error throughout. Among the private methods, STR performs best across all three settings: its error decreases as either \(\varepsilon\) or \(n\) increases, and it consistently improves over WPE. The advantage of STR is especially clear in the last regime, where larger values of \(a\) place more observations near \(x_0\), making the local kernel estimate more stable and increasing the probability of a successful private release.

\section{Wine quality linear regression}
We consider the Wine Quality dataset from the UCI Machine Learning Repository, originally collected for modeling Portuguese Vinho Verde wine quality from physicochemical measurements
\citep{cortez2009modeling}. We take the quality score as the response and selected physicochemical variables as
predictors, including alcohol, volatile acidity, density, and pH. 
The original dataset contains $6{,}497$ wine samples. We take a $20\%$ training sample and a held-out $80\%$ testing sample in each replicate. 

\begin{figure}
    \centering
    \includegraphics[width=0.9\linewidth]{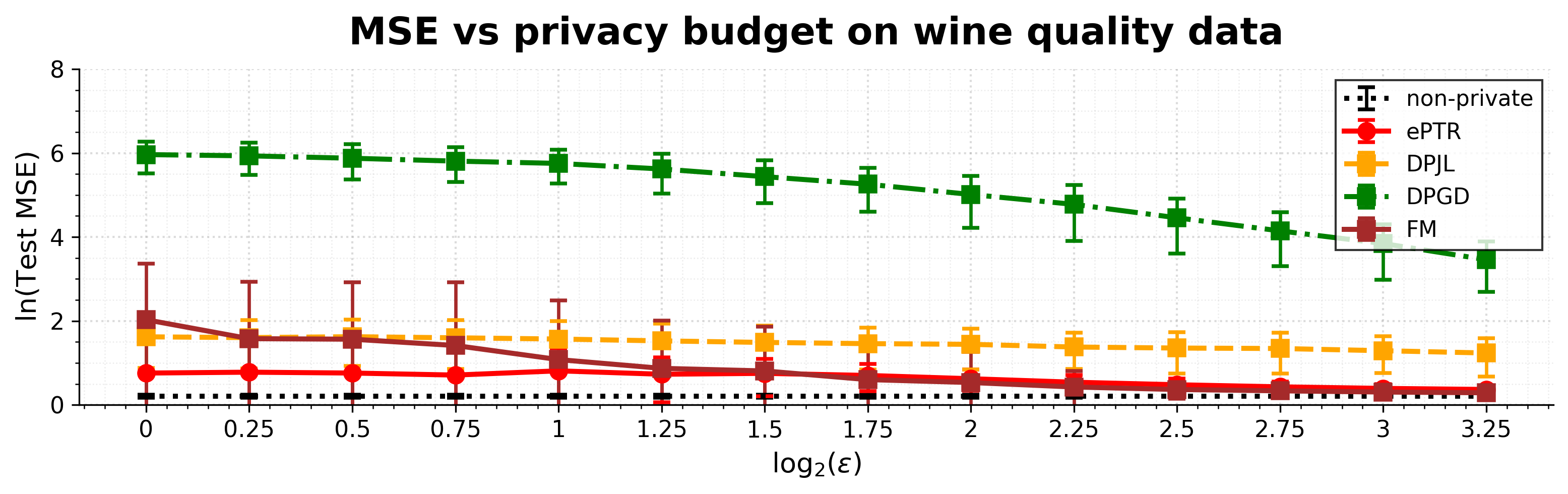}
    \caption{Logarithm of test MSE versus $\log_2(\varepsilon)$ for linear regression on wine quality data.}
    \label{fig:wine}
\end{figure}

We apply STR-OLS and the private methods discussed in Section \ref{subsec:simu_linear}. For STR-OLS, we set $c_0=0.062$, $R_\theta=4$,
$R_x=2$, and $R_y=3$. These bounds apply after
standardizing the covariates using the training sample and transforming the response as $(Y-5)/5$. Meanwhile, we take $\bot = 5$ as data-independent model prediction output, since the quality score has a publicly known range $[0,10]$. 
The test MSEs over 50 replicates are shown in Figure~\ref{fig:wine}. 
Most methods have a relatively large MSE in the strong privacy regime $\varepsilon \leq 2$, but this error decreases with an increasing privacy budget $\varepsilon$. 
In contrast, STR performs best throughout, even in the strong privacy regime. As $\varepsilon$ increases, its test MSE gradually decreases and moves toward the non-private benchmark.

\section{Proof for the STR scheme}
\subsection{Proof of Theorem~\ref{thm:ePTRissafe}}
\begin{proof}
Let the test result $I(\calX)=1_{\thetatilde_\gamma=\thetatilde}$. Our first step is showing that $I(\calX)$ is $(\frac12\varepsilon,0)$-DP. 
We note that the following holds for all $\gamma,\varepsilon\geq 0$.  
$$\frac{e^{\frac12\varepsilon \gamma}}{e^{\frac12\varepsilon \gamma}+1}\leq \frac{e^{\frac12\varepsilon(\gamma-1)}e^{\frac12\varepsilon}}{e^{\frac12\varepsilon (\gamma-1)}+1},\quad \frac{1}{e^{\frac12\varepsilon \gamma}+1}\leq \frac{e^{\frac12\varepsilon}}{e^{\frac12\varepsilon (\gamma+1)}+1}.$$
Recall that $|\gamma(\calX)-\gamma(\calX')|\leq 1$.
So for any $i\in\{0,1\}$
\[
P(I(\calX')=i) e^{-\frac12\varepsilon}\leq P(I(\calX)=i)\leq P(I(\calX')=i) e^{\frac12\varepsilon}.
\]
Meanwhile, for any set $\calB$, we decompose 
\begin{align*}
P(\thetatilde_\gamma(\calX)\in \calB)&=P(\thetatilde_\gamma (\calX)\in \calB, I(\calX)=1)+P(\thetatilde_\gamma (\calX)\in \calB, I(\calX)=0)\\
&=P(\thetatilde (\calX)\in \calB)P(I(\calX)=1)+ \pi_\bot(\calB)P(I(\calX)=0),
\end{align*}
where $\pi_\bot(\calB)$ describes the distribution of the $\bot$. 

When $\gamma (\calX)>0$, since $\{\gamma>0\}\subseteq sub_\alpha$, 
we have $\|\thetahat(\calX)-\thetahat(\calX')\|\leq \alpha$. Then
by the Gaussian DP mechanism, with $\tilde\theta=\tilde\theta_\alpha=\hat\theta+\frac{2\alpha}{\varepsilon}\sqrt{2\log(1.25/\delta)}\,\zeta$,  we have 
\[
P(\thetatilde_\alpha(\calX)\in \calB)\leq P(\thetatilde_\alpha(\calX')\in \calB) e^{\frac12\varepsilon}+\delta,
\]
so we have
\begin{align*}
&P(\thetatilde_\gamma(\calX)\in \calB)\\
&=P(\thetatilde_\alpha(\calX)\in \calB)P(I(\calX)=1)+\pi_\bot(\calB)P(I(\calX)=0)\\
&\leq e^{\frac12\varepsilon} P(\thetatilde_\alpha(\calX')\in \calB) e^{\frac12\varepsilon} P(I(\calX')=1)+P(I(\calX)=1)\delta+\pi_\bot(\calB) P(I(\calX')=0)e^{\frac12\varepsilon}\\
&\leq  e^{\varepsilon} \left(P(\thetatilde_\alpha(\calX')\in \calB)P(I(\calX')=1)+\pi_\bot(\calB) P(I(\calX')=0)\right)+\delta
\\
&\leq e^{\varepsilon}P(\thetatilde_\gamma (\calX')\in \calB)+\delta. 
\end{align*}
Suppose $\gamma(\calX)= 0$, then $\gamma(\calX')\leq 1$, and
\[
P(I(\calX)=1)=\frac{e^{-\frac12 M\varepsilon}}{e^{-\frac12 M\varepsilon}+1}\leq\delta e^{-\frac12\varepsilon}, P(I(\calX')=1)\leq \frac{e^{-\frac12(M-1)\varepsilon}}{e^{-\frac12(M-1)\varepsilon}+1}\leq\delta. 
\]
We have 
\begin{align*}
P(\thetatilde_\gamma(\calX)\in \calB)
&=P(\thetatilde(\calX)\in \calB)P(I(\calX)=1)+ \pi_\bot(\calB)P(I(\calX)=0)\leq \delta+\pi_\bot(\calB). 
\end{align*}
Likewise
\begin{align*}
P(\thetatilde_\gamma(\calX')\in \calB)
&=P(\thetatilde(\calX')\in \calB)P(I(\calX')=1)+ \pi_\bot(\calB)P(I(\calX')=0)\\
&\geq \frac{e^{\frac12(M-1)\varepsilon}}{e^{\frac12(M-1)\varepsilon}+1}\pi_\bot(\calB)
\end{align*}
Then because $e^{\varepsilon}\geq 1+\varepsilon$ while $e^{\frac12 (M-1)\varepsilon}\geq \frac{1}{\varepsilon}$, 
\[
e^{\varepsilon}\frac{e^{\frac12(M-1)\varepsilon}}{e^{\frac12(M-1)\varepsilon}+1}
=e^{\varepsilon}\frac{1}{e^{-\frac12(M-1)\varepsilon}+1}\geq (1+\varepsilon)\frac{1}{\varepsilon+1}\geq 1. 
\]
\[
P(\thetatilde_\gamma (\calX)\in \calB)\leq \delta+\pi_\bot(\calB)\leq 
e^{\varepsilon}\frac{e^{\frac12(M-1)\varepsilon}}{e^{\frac12(M-1)\varepsilon}+1} \pi_\bot(\calB)+\delta\leq e^{\varepsilon}P(\thetatilde_\gamma (\calX')\in \calB)+\delta. 
\]
\end{proof}

\subsection{Proof of Theorem~\ref{thm:ePTRperformance}}
\begin{proof}
Let the DP test result $I(\calX)=1_{\thetatilde_\gamma=\thetatilde_\alpha}$.
We simply decompose 
\begin{align*}
\E_\theta[\|\thetatilde_\gamma(\calX)-\theta\|^2]&=
\E_\theta[\|\thetatilde_\gamma(\calX)-\theta\|^2I(\calX)]+
\E_\theta[\|\thetatilde_\gamma(\calX)-\theta\|^2(1-I(\calX))]\\
&=\E_\theta[\|\thetatilde_\alpha-\theta\|^2I(\calX)]+
\E_\theta[\|\bot-\theta\|^2(1-I(\calX))]\\
&\leq \E_\theta[\|\thetatilde_\alpha-\theta\|^2]+C_\bot P_\theta(I(\calX)=0)\\
&= \text{(I)}+\text{(II)}+C_\bot P_\theta(I(\calX)=0).
\end{align*}
Finally we note that $P_\theta(I(\calX)=0|\gamma(\calX)\geq 2M)\leq e^{-\frac12M\varepsilon}\leq \delta$. So
\begin{align*}
P_\theta(I(\calX)=0)&\leq P_\theta(I(\calX)=0|\gamma(\calX)\geq 2M)P_\theta(\gamma(\calX)\geq 2M)+
 P_\theta(\gamma(\calX)< 2M)\\
&\leq \delta+P_\theta(\gamma(\calX)< 2M). 
\end{align*}
This leads to (III). 
\end{proof}

\section{Proof for private Bayes classifier}
\subsection{Proof of Theorem~\ref{thm:BayePTRDP}}
\begin{proof}
    First, we verify \eqref{eqn:bayesperturb}. Note that if a $(x_i,y_i)$ is replaced with $(x'_i,y'_i)$. There are four possibilities
\begin{enumerate}
     \item $y_i=k=y'_i$. The resulting change for $\mhat_k$ will be 
    $\frac{x_i'-x_i}{|I_k|}$. Its norm is at most $(R_x+\max_{j\in I_k}\|x_j\|)/|I_k|$, attained when $\|x_i\|=\max_{j\in I_k}\|x_j\|$ and $x_i'=-R_xx_i/\|x_i\|$.
    \item $y_i=k\neq y'_i=j$. The resulting change will be 
    \[\Big\|\mhat_k-\frac{\mhat_k |I_k|-x_i}{|I_k|-1}\Big\|=\frac{\|x_i-\mhat_k\|}{|I_k|-1}\leq \frac{\sum_{l\in I_k\setminus\{i\}}\|x_l\|}{(|I_k|-1)|I_k|}+\frac{\|x_i\|}{|I_k|}.
    \]
    \item $y_i=j\neq k=y'_i$. The resulting change will be 
    $\mhat_k-\frac{\mhat_k |I_k|+x'_i}{|I_k|+1}=\frac{\mhat_k-x'_i}{|I_k|+1}$. 
    \item $y_i\neq k\neq y'_i$. There will be no changes to $\mhat_k$. 
\end{enumerate}
Maximizing over these gives us \eqref{eqn:bayesperturb}.

Since Algorithm~\ref{alg:ePTRBC} uses $\Pi_{R_x}(x_i)$ in place of $x_i$ and $\|\Pi_{R_x}(x_i)\|\leq R_x$, \eqref{eqn:bayesperturb} holds with every $\|x_j\|$ bounded by $R_x$; and if $\gamma(\calX)>0$, then $|I_k|-1>c_0n$ for all $k$. Hence
\[
\Delta_k(\calX)\leq \max\{\frac{2R_x}{c_0n},\frac{2R_x}{c_0n}, \frac{2R_x}{c_0n+1}\}\leq \frac{2R_x}{c_0n}.
\]
Note also that in any of these four situations, only at most two $\mhat_i$ and $\hat{\mu}_i$ will be perturbed. 
The overall perturbation is 
\[
\big\|(\mhat_1,\ldots,\mhat_K,\hat{\mu}_1,\ldots,\hat{\mu}_K)(\calX)-(\mhat_1,\ldots,\mhat_K,\hat{\mu}_1,\ldots,\hat{\mu}_K)(\calX')\big\|\leq \frac{2}{n}\sqrt{\frac{2R^2_x}{c^2_0 }+2}=:\alpha. 
\]

In other words, $\{\gamma>0\}\subseteq sub_{\alpha}$. Finally, because $||I_k(\calX)|-|I_k(\calX')||\leq 1$, $\gamma$ is $1$-Lipschitz with respect to $D_H$. Theorem \ref{thm:ePTRissafe} shows that STR-Bayes is $(\varepsilon,\delta)$-DP. 

\end{proof}

\subsection{Proof of Theorem~\ref{thm:BayePTR}}
\begin{proof}
Write $\theta=(\mu_1,\ldots,\mu_K,m_1,\ldots,m_K)$ and $\thetahat=(\hat\mu_1,\ldots,\hat\mu_K,\mhat_1,\ldots,\mhat_K)$, computed from the projected data, with $\mhat_k:=0$ if $I_k=\emptyset$. Assume $\log n\geq 4$.

\emph{Release probability.} Since $\delta<n^{-3}$ and $8\log(1/\delta)/\varepsilon\leq c_0n-4$, we have $\varepsilon\geq 24\log n/(c_0n)\geq\delta$, so $M=1+\frac2\varepsilon\log(1/\delta)$ and
\[
2M=2+\frac{4}{\varepsilon}\log(1/\delta)\leq \frac{c_0n}{2}.
\]
Hence $\gamma(\calX)\geq 2M$ whenever $\min_k|I_k|\geq \frac32 c_0n+1$. Since $|I_k|\sim\mathrm{Bin}(n,\mu_k)$ with $\mu_k>2c_0$, Bernstein's inequality gives, for $n$ sufficiently large,
\[
P\Big(|I_k|<\tfrac32c_0n+1\Big)
\leq \exp\left(-\frac{(\frac12c_0n-1)^2}{2n+\frac23c_0n}\right)
\leq \exp\big(-\tfrac{1}{14}c_0^2n\big).
\]
So $P(\gamma(\calX)<2M)\leq K\exp(-\frac1{14}c_0^2n)\leq n^{-3}$ for $n$ sufficiently large.

\emph{Error of $\thetahat$.} Write $x_j=m_{y_j}+\xi_j$ with $\xi_j\sim\mathcal N(0,I_p)$. Since $\|m_k\|\leq C_m\sqrt p$ with $C_m\geq1$, $R_x-\|m_k\|\geq \sqrt p+3\sqrt{p\log n}$, so
\[
P\Big(\max_i\|x_i\|>R_x\Big)\leq n\exp\Big(-\frac92p\log n\Big)\leq n^{-7/2}.
\]
Let $A=\{\max_i\|x_i\|\leq R_x,\ \min_k|I_k|\geq \frac32c_0n+1\}$ and $A_y=\{\min_k|I_k|\geq \frac32c_0n+1\}\supseteq A$, so $P(A^c)=O(n^{-3})$. On $A$ the projection does not change the data, $\mhat_k=m_k+\frac{1}{|I_k|}\sum_{j\in I_k}\xi_j$. Since $A_y$ depends only on $y_{[n]}$ and $\xi_{[n]}$ is independent of $y_{[n]}$,
\[
\E\big[\|\mhat_k-m_k\|^21_A\big]
\leq \E\left[\E\left[\left\|\frac{1}{|I_k|}\sum_{j\in I_k}\xi_j\right\|^2\bigg| y_{[n]}\right]1_{A_y}\right]
=\E\left[\frac{p}{|I_k|}1_{A_y}\right]\leq \frac{p}{c_0n}.
\]
Together with $\E[|\hat\mu_k-\mu_k|^2]=\mu_k(1-\mu_k)/n$,
\[
\E[\|\thetahat-\theta\|^21_A]=O\Big(\frac{p}{n}\Big).
\]
On $A^c$, $\|\mhat_k\|\leq R_x$, so $\E[\|\thetahat-\theta\|^21_{A^c}]=O(R_x^2n^{-3})=o(p/n)$.

\emph{STR release.} The released vector has dimension $K(p+1)$ and $\alpha=\frac2n\sqrt{\frac{2R_x^2}{c_0^2}+2}$, so $\alpha^2=O(p\log n/n^2)$. With a fixed $\bot=0$, $C_\bot=O(p)$. Theorem~\ref{thm:ePTRperformance} gives, for the STR output $(\tilde\mu_{[K]},\tilde m_{[K]})$ before step 3 of Algorithm~\ref{alg:ePTRBC},
\[
\E\Big[\sum_{k=1}^K\big(|\mu_k-\tilde\mu_k|^2+\|m_k-\tilde m_k\|^2\big)\Big]
=O\Big(\frac{p}{n}+\frac{p^2\log n\log(1/\delta)}{\varepsilon^2n^2}\Big).
\]

\emph{Renormalisation.} Let $v_k=\tilde\mu_k\vee c_0$ and $S=\sum_{j=1}^Kv_j\geq Kc_0$. Since $\mu_k>c_0$, $|v_k-\mu_k|\leq|\tilde\mu_k-\mu_k|$, and $|S-1|=|\sum_j(v_j-\mu_j)|$. Hence
\[
\sum_{k=1}^K\Big|\frac{v_k}{S}-\mu_k\Big|^2
\leq \frac{2}{S^2}\sum_{k=1}^K\big(|v_k-\mu_k|^2+\mu_k^2|S-1|^2\big)
\leq \frac{2(1+K)}{K^2c_0^2}\sum_{k=1}^K|\tilde\mu_k-\mu_k|^2,
\]
so step 3 changes the bound by a constant factor only.
\end{proof}
\section{Proof for private linear regression}
\subsection{Proof of Theorem~\ref{thm:LRDPcheck}}
\begin{proof}

Without loss of generality, we assume $\|\thetahat(\calX)\|\leq \|\thetahat(\calX')\|$. Else we swap their roles in below. 
We start by verification of the local sensitivity. Because all $\|x_i\|\leq R_x, |y_i|\leq R_xR_\theta$ through the projection, suppose $\calX'$ replaces $(x_i,y_i)$ with $(x'_i,y'_i)$, we have  
\begin{align*}
&\|\thetahat(\calX)-\thetahat(\calX')\|\\
&=\|(X'(X')\ic)^{-1}X'Y'-(XX\ic)^{-1}XY\|\\
&\leq\|(X'(X')\ic)^{-1}(X'Y'-XY)\|+\|((X'(X')\ic)^{-1}-(XX\ic)^{-1})XY\|\\
&=\|(X'(X')\ic)^{-1}(y_i'x_i'-y_ix_i)\|+\|(X'(X')\ic)^{-1}(XX\ic-X'(X')\ic)(XX\ic)^{-1}XY\|\\
&\leq \|(X'(X')\ic)^{-1}\|\|y'_ix'_i-y_ix_i\|+\|(X'(X')\ic)^{-1}\|\|x_ix_i\ic-x'_i(x'_i)\ic\|\|\thetahat(\calX)\|\\
&\leq 2R_x^2\|(X'(X')\ic)^{-1}\|(\|\thetahat(\calX)\|+R_\theta)
\end{align*}
Then by Lemma \ref{lem:rperturb}
\[
\|\Pi_{R_\theta}(\thetahat(\calX))-
\Pi_{R_\theta}(\thetahat(\calX'))\|
\leq 2R_x^2\|(X'(X')\ic)^{-1}\|\frac{\|\thetahat\|+R_\theta}{(\|\thetahat\|/R_\theta)\vee 1}
\leq 4R_x^2R_\theta\|(X'(X')\ic)^{-1}\| 
\]

Next, by eigenvalue perturbation, 
\[
|\lambda_{p}(XX\ic)-\lambda_{p}(X'(X')\ic)|
\leq \|x_ix_i\ic-x_i'(x_i')\ic\|
\leq 2R_x^2.
\]
So if $\gamma(\calX)>0$, then  $\lambda_{p}(X'(X')\ic)\geq c_0 n$, and 
$\Delta_{\thetahat}(\calX)\leq \frac{4R_x^2R_\theta}{c_0n}$. 
Consequently, $\{\gamma>0\}\subseteq sub_\alpha$. 

From above, we also see 
\[
|\gamma(\calX)-\gamma(\calX')|
\leq \frac{1}{2R_x^2}|\lambda_p(XX\ic)-\lambda_p(X'(X')\ic)|\leq 1. 
\]
Hence $\gamma$ is an $\alpha$-safety score, and Theorem \ref{thm:ePTRissafe} shows that STR-OLS is $(\varepsilon,\delta)$-DP.
    
\end{proof}
\subsection{Proof of Theorem~\ref{thm:LRDP}}
\begin{proof}
Throughout, $n$ is sufficiently large, $\log n\geq 4$, and $c_0\leq 1/(4C_x^2)$ as set before Theorem~\ref{thm:LRDP}. Since $C_x^{-2}\leq\lambda_p(\Sigma)\leq\lambda_1(\Sigma)\leq C_x^2$, we have $C_x\geq 1$.

\emph{Step 1: the data are not clipped with high probability.}
By Gaussian concentration of the norm, $P(\|\Sigma^{-1/2}x_i\|\geq \sqrt p+t)\leq e^{-t^2/2}$ for $t\geq 0$, and $\|x_i\|\leq C_x\|\Sigma^{-1/2}x_i\|$. Since
\[
\frac{R_x}{C_x}-\sqrt{p}=\sqrt{p}\,\big(2\sqrt{\log n}-1\big)\geq \frac32\sqrt{p\log n}
\]
and $p\geq 4$, a union bound gives
\[
P\Big(\max_{i\in[n]}\|x_i\|> R_x\Big)\leq n\exp\Big(-\frac98 p\log n\Big)\leq n^{-7/2}.
\]
Next, $y_i\sim\mathcal{N}(0,\theta\ic\Sigma\theta+\sigma_\xi^2)$. By Assumption~\ref{aspt:LR} and $p\geq 4$,
\[
\theta\ic\Sigma\theta+\sigma_\xi^2\leq C_x^2R_\theta^2+\frac14C_x^2R_\theta^2\,p\leq \frac12 C_x^2R_\theta^2\,p.
\]
Using $P(|Z|>t)\leq 2e^{-t^2/2}$ for $Z\sim\mathcal{N}(0,1)$ and $R_x^2=4C_x^2p\log n$,
\[
P\Big(\max_{i\in[n]}|y_i|> R_xR_\theta\Big)\leq 2n\exp\Big(-\frac{R_x^2R_\theta^2}{C_x^2R_\theta^2\,p}\Big)=2n^{-3}.
\]

\emph{Step 2: the design is well conditioned.}
Let $E=\Sigma^{-1/2}X\in\reals^{p\times n}$, whose entries are i.i.d.\ $\mathcal{N}(0,1)$. By \citet[Corollary 5.35]{vershynin2010introduction}, with probability at least $1-2n^{-4}$,
\[
\sqrt{\lambda_p(EE\ic)}\geq \sqrt n-\sqrt p-\sqrt{8\log n}\geq \frac{\sqrt3}{2}\sqrt n,
\]
where the last step uses $p\leq n/(64C_x^4\log n)$. Hence, with probability at least $1-2n^{-4}$,
\[
\lambda_p(XX\ic)\geq \lambda_p(\Sigma)\lambda_p(EE\ic)\geq \frac34 n C_x^{-2}.
\]

\emph{Step 3: the good event.} Let
\[
A=\Big\{\lambda_p(XX\ic)\geq \frac34 nC_x^{-2},\ \|x_i\|\leq R_x,\ |y_i|\leq R_x R_\theta,\ \forall i\in[n]\Big\}.
\]
By Steps 1 and 2, $P(A^c)\leq 2n^{-4}+n^{-7/2}+2n^{-3}\leq 3n^{-3}$.

\emph{Step 4: term (I).} On $A$, the projections of $x_i$ and $y_i$ in Algorithm~\ref{alg:ePTROLS} are inactive, so the least-squares solution equals $\theta+(XX\ic)^{-1}X\xi$; moreover $\Pi_{R_\theta}$ is non-expansive and $\Pi_{R_\theta}(\theta)=\theta$. On $A^c$, $\|\Pi_{R_\theta}(\thetahat)-\theta\|\leq 2R_\theta$. Let $A_X=\{\lambda_p(XX\ic)\geq \frac34 nC_x^{-2}\}\supseteq A$. Since $A_X$ depends only on $X$ and $\xi$ is independent of $X$,
\begin{align*}
\E[\|\Pi_{R_\theta}(\thetahat)-\theta\|^2]
&\leq \E[\|(XX\ic)^{-1}X\xi\|^2 1_{A_X}]+4R_\theta^2P(A^c)\\
&= \sigma_\xi^2\,\E[\text{tr}((XX\ic)^{-1})1_{A_X}]+4R_\theta^2P(A^c)\\
&\leq \frac{4p \sigma_\xi^2 C_x^2}{3n}+12n^{-3}R_\theta^2.
\end{align*}

\emph{Step 5: the release probability.} On $A$, using $c_0\leq 1/(4C_x^2)$ and $R_x^2=4C_x^2p\log n$,
\[
\gamma(\calX)=\frac{1}{2R_x^2}\big(\lambda_p(XX\ic)-c_0n-2R_x^2\big)_+
\geq \frac{\frac12 nC_x^{-2}}{8C_x^2p\log n}-1
=\frac{n}{16C_x^4p\log n}-1
\geq\frac{n}{32C_x^4p\log n},
\]
where the last step uses $n/(32C_x^4p\log n)\geq 1$. The condition of the theorem gives $\varepsilon\geq 128pC_x^4\log n\log(1/\delta)/n\geq\delta$, so $M=1+\frac2\varepsilon\log(1/\delta)$, and the condition is equivalent to $\frac{n}{32C_x^4p\log n}\geq 2M$. Hence $\gamma(\calX)\geq 2M$ on $A$, and $P(\gamma(\calX)<2M)\leq P(A^c)\leq 3n^{-3}$.

\emph{Step 6: conclusion.} Apply Theorem~\ref{thm:ePTRperformance} with $\bot=0$, so that $C_\bot=R_\theta^2$. Term (I) is bounded in Step 4. For term (II), $\alpha=\frac{4R_x^2R_\theta}{c_0n}=\frac{16C_x^2R_\theta\, p\log n}{c_0n}$, so
\[
\frac{8\alpha^2p}{\varepsilon^2}\log\Big(\frac{1.25}{\delta}\Big)\leq C\,R_\theta^2\frac{p^3(\log n)^2\log(1/\delta)}{\varepsilon^2n^2},
\]
with $C$ depending only on $C_x$ and $c_0$. Term (III) is at most $R_\theta^2(3n^{-3}+\delta)\leq 4R_\theta^2n^{-3}$. Since $\varepsilon<2$ and $\log(1/\delta)\geq 3\log n$, the terms $12R_\theta^2n^{-3}$ and $4R_\theta^2n^{-3}$ are bounded by the second term in the theorem. This gives the claimed bound.
\end{proof}

\subsection{Proof of Lemma~\ref{lem:rperturb}}

\begin{supplem}
\label{lem:rperturb}
For two vectors $v,v'$ and a scalar $a>0$, the following hold:
\[
\left\|\frac{v}{\|v\|\vee a}-\frac{v'}{\|v'\|\vee a}\right\|\leq (\frac{\left\|v-v'\right\|}{(\|v'\|\wedge \|v\|)\vee a}). 
\]
\end{supplem}
\begin{proof}
Without loss of generality, we assume $\|v\|\leq \|v'\|$. 
Note that we can always divide all vectors by $a$, which reduces the problem to showing 
\[
\left\|\frac{v}{\|v\|\vee 1}-\frac{v'}{\|v'\|\vee 1}\right\|\leq (\frac{\left\|v-v'\right\|}{\|v\|\vee 1})
\]
There are three scenarios. In the first scenario, $\|v\|\leq\|v'\|\leq 1$, then the claim is an identity. In the second scenario, 
$\|v\|\leq 1\leq \|v'\|$. Then we want to show
\[
\left\|v-\frac{v'}{\|v'\|}\right\|\leq \left\|v-v'\right\|,
\]
in other words, the projection of point $v'$ on unit ball is closer to $v$, a point in the unit ball, than $v'$ is. This is a well known property due to the convexity of the unit ball. 

Finally, if $1<r=\|v\|\leq \|v'\|=b r$ with $b\geq 1$. Suppose the angle between $v$ and $v'$ is $\varphi$, then the left side is 
\[
\|\frac{v}{\|v\|}-\frac{v'}{\|v'\|}\|^2= 2-2\cos \varphi =4\sin (\frac{\varphi}{2})^2.
\]
The right hand side is 
\[
\frac{\|v-v'\|^2}{\|v\|^2}= 1-2\cos \varphi b+ b^2=:g(b).
\]
Since $\dot{g}(b)=2b-2\cos\varphi\geq 0$ when $b\geq 1$, we find that $g(b)\geq g(1)=$ left hand side. 

In conclusion, we have our claim.
\end{proof}

\section{Proof for private kernel regression}

\subsection{Technical details of kernel regression}
\begin{supplem}
\label{lem:outlier}
If $K_\sigma(x,y)$ is a regular radial basis kernel with bandwidth $\sigma$, then for any function $|g|\leq C_g$ and density $\mu\leq C_\mu$ the following holds for $r=2C_K\sigma\sqrt{\log n}$ and $n$ sufficiently large
\[
\int_{\|u\|>\frac{r}{\sigma}} |K(u)g(x+\sigma u)\mu(x+\sigma u)|du\leq n^{-3},
\]
\[
\left|
\int_{B^c_{r}(x)} K_\sigma(x,y)g(y)\mu(y)dy\right|=O(n^{-3}\sigma^{-d}),
\]
\[
\left|
\sum_{i\in [n]}1_{x_i\in B^c_r(x)} K_\sigma(x,x_i)g(x_i)\right|=O(n^{-3}\sigma^{-d}).
\]
\end{supplem}
\begin{supplem}
\label{lem:ind}
Let $F_i=K_{\sigma}(x_0,x_i)f(x_i)$, then $|F_i|\leq C_fC_K\sigma^{-d}$ a.s., while 
\[
\left|\E[F_i]-f(x_0)\mu(x_0)\right|=O(n^{-3}\sigma^{-d}+\sigma^{s}),
\]
\[
\text{var}[F_i]\leq O(V_dr^d\sigma^{-2d}+n^{-3}\sigma^{-2d}).
\]
\end{supplem}

\begin{supppro}
\label{pro:NW}
Suppose $f(x)\mu(x)$ is $\mathcal{C}^s$ at $x_0$ and $n\sigma^d\geq 1$,
then the following estimate holds with probability at least $1-n^{-5}$:
\[
\left|\frac1n\sum_{i\in [n]} K_\sigma(x_0,x_i)f(x_i)- f(x_0)\mu(x_0)\right|=O\left(\sigma^s+\frac{(\log n)^{1+d/2}}{\sqrt{n\sigma^d}}\right).
\]
\end{supppro}

\begin{supplem}
\label{lem:buildK}
Let $\phi_\sigma(\|v\|)=\frac{1}{\sqrt{2\sigma^2\pi}^d}\exp(-\frac{\|v\|^2}{2\sigma^2})$ be the Gaussian density with mean zero and variance $\sigma^2 I_d$, and write $\phi_i:=\phi_\sigma$ with $\sigma=i$, i.e.\ the Gaussian density with standard deviation $i$. Then given any positive integer $s$, a regular radial basis kernel of order $s$ can be written as
\[
K(\|v\|)=\sum_{i=1}^s a_{i} \phi_i(\|v\|)
\]
for some $a_i\in \reals$.
\end{supplem}
\subsection{Proof of Theorem~\ref{thm:KRDP}}
\begin{proof}

Without loss of generality, we assume $|\thetahat(\calX)|\leq|\thetahat(\calX')|$, else we swap the two in the derivation below.
We note that 
\begin{align*}
|\thetahat(\calX)-\thetahat(\calX')|&
=\left|\frac{\sum_{j\in[n]}K_{\sigma}(x,x_j)y_j}{d_\sigma(x,\calX)}-\frac{\sum_{j\in[n]}K_{\sigma}(x,x'_j)y'_j}{d_\sigma(x,\calX')}\right|\\
&\leq \frac{|K_\sigma(x,x_i)y_i-K_\sigma(x,x'_i)y'_i|}{d_{\sigma}(x,\calX')}+
\frac{|K_\sigma(x,x_i)-K_\sigma(x,x'_i)|}{d_{\sigma}(x,\calX')}|\thetahat(\calX)|\\
&\leq \frac{2R_fC_K+2C_K|\thetahat(\calX)|}{\sigma^d d_{\sigma}(x,\calX')}. 
\end{align*}

Therefore, Lemma \ref{lem:rperturb} leads to 
\[
|\Pi_{R_f}\thetahat(\calX)-\Pi_{R_f}\thetahat(\calX')|
\leq \frac{R_f(2R_fC_K+2C_K|\thetahat(\calX)|)}{\sigma^d d_{\sigma}(x,\calX')(|\thetahat(\calX)|\vee R_f )}\leq \frac{4R_fC_K}{\sigma^d d_{\sigma}(x,\calX')}.
\]
So if $\gamma(\calX)>0$, we have $d_{\sigma}(x,\calX')>c_0 n$. This leads to 
$\{\gamma>0\}\subset sub_{\alpha}$, with $\alpha=\frac{4R_fC_K}{c_0 n\sigma^d}$.
Finally we apply Theorem \ref{thm:ePTRissafe}.

\end{proof}

\subsection{Proof of Theorem~\ref{thm:KR} and its corollary}
\begin{proof}
With our choice of $R_f$, it is easy to see that with high probability, the projection will not change the data points. Indeed, $|y_i|\leq C_f+|\xi_i|$ with $\xi_i\sim\mathcal{N}(0,\sigma_\xi^2)$, $\sigma_\xi\leq C_f$, so $P(|\xi_i|>C_f\sqrt{8\log n})\leq 2n^{-4}$, and a union bound gives $\max_i|y_i|\leq R_f=C_f(1+\sqrt{8\log n})$ with probability at least $1-2n^{-3}$.

Using Proposition \ref{pro:NW}, we find that with high probability $1-n^{-5}$
\[
\left|\frac1n\sum_{i\in [n]} K_\sigma(x_0,x_i)f(x_i)- f(x_0)\mu(x_0)\right|=O(\sigma^s+\frac{(\log n)^{1+d/2}}{\sqrt{n\sigma^d}})
\]
\[
\left|\frac1n\sum_{i\in [n]} K_\sigma(x_0,x_i)- \mu(x_0)\right|=O(\sigma^s+\frac{(\log n)^{1+d/2}}{\sqrt{n\sigma^d}})
\]
Meanwhile, conditioned on $x_{[n]}$, the noise part satisfies
\[
\frac{\frac1n\sum_{i\in [n]} K_\sigma(x_0,x_i)\xi_i}{\frac1n\sum_{i\in [n]} K_\sigma(x_0,x_i)}\sim \mathcal{N}(0,\sigma_\xi^2 \tau^2),
\qquad
\tau^2=\frac{\sum_{i\in [n]} K_\sigma(x_0,x_i)^2}{d_{\sigma}(x_0,\calX)^2}. 
\]
Note further that by Lemma \ref{lem:outlier}
\begin{align*}
\sum_{i\in [n]} K_\sigma(x_0,x_i)^2&\leq \sum_{i\in [n]} 1_{x_i\in B_r(x_0)}K_\sigma(x_0,x_i)^2+C_K^2 n^{-5}\sigma^{-2d}\\
&\leq \Big(\sum_{i\in [n]} 1_{x_i\in B_r(x_0)}+n^{-5}\Big)C_K^2 \sigma^{-2d}\\
&\leq \big(2 n\mu(x_0)V_d r^{d}+ n^{-5}\big)C_K^2\sigma^{-2d}
\end{align*}
with high probability. Since $d_\sigma(x_0,\calX)\geq\frac12n\mu(x_0)$ with high probability,
\[
\tau^2\leq \bar\tau^2:=\frac{9V_dC_K^2\,r^d \sigma^{-2d}}{n\mu(x_0)}.
\]
Therefore
\[
P\left(\left|\frac{\frac1n\sum_{i\in [n]} K_\sigma(x_0,x_i)\xi_i}{\frac1n\sum_{i\in [n]} K_\sigma(x_0,x_i)}\right|>\sigma_\xi\sqrt{8\log n\,\bar\tau^2}\right)\leq 2n^{-4}+P(\tau^2>\bar\tau^2)=O(n^{-4}). 
\]
Putting these together, and the fact $\Pi_{R_f}$ will confine $\fhat$ into a bounded set, we have 
\[
\E[|\fhat(x_0)-f(x_0)|^2]=O\left(\sigma^{2s}+\frac{(\log n)^{2+d}}{n \sigma^d}\right). 
\]

The last step is checking that $P(\gamma(\calX)<2M)$ is negligible. This is true because by Proposition \ref{pro:NW}, with probability at least $1-n^{-5}$
\[
\left|\sum_{i\in [n]} K_\sigma(x_0,x_i)- n\mu(x_0)\right|=O(n\sigma^s+n\frac{(\log n)^{1+d/2}}{\sqrt{n\sigma^d}}).
\]
The right-hand side is $o(n)$ under the conditions of the theorem.
So if $\mu(x_0)>2c_0+\frac{4}{n\sigma^d}C_K$, then $c_0n+2\sigma^{-d}C_K<\frac12 n\mu(x_0)$, and for $n$ sufficiently large, with probability at least $1-n^{-3}$,
\[
\gamma(\calX)\geq \frac{1}{2\sigma^{-d}C_K}\left(\sum_{i\in [n]} K_\sigma(x_0,x_i)- c_0 n-2\sigma^{-d} C_K\right)\geq
\frac{\sigma^d}{2C_K}\cdot\frac14 n\mu(x_0)=\frac{n\sigma^d\mu(x_0)}{8C_K}.
\]
Since $\delta\leq\varepsilon$, the threshold in Algorithm~\ref{alg:ePTR} is $M=1+\frac{2}{\varepsilon}\log(1/\delta)$. By the condition $n\sigma^d\mu(x_0)\geq 16C_K\big(1+\frac{2}{\varepsilon}\log(1/\delta)\big)$, we have $\gamma(\calX)\geq 2M$, so $P(\gamma(\calX)<2M)\leq n^{-3}$.

Finally, we apply Theorem~\ref{thm:ePTRperformance} with $p=1$. Term (I) is bounded above. For term (II), $\alpha=\frac{4R_fC_K}{c_0n\sigma^d}$ and $R_f^2=O(\log n)$, so
\[
\frac{8\alpha^2}{\varepsilon^2}\log\Big(\frac{1.25}{\delta}\Big)=O\left(\frac{\log n\,\log(1/\delta)}{\varepsilon^2(n\sigma^d)^2}\right).
\]
For term (III), $C_\bot=O(1)$ when the no-reply output $\bot$ is a fixed bounded value (e.g.\ $0$), and $P(\gamma(\calX)<2M)+\delta\leq 2n^{-3}$, which is negligible. This leads to our final estimate.
\end{proof}

\subsection{Proof of Lemma~\ref{lem:outlier}}
\begin{proof}
For $\|u\|>r/\sigma$, we have $\|u\|^2/(2C_K)\geq 2C_K\log n\geq 4\log n$ since $C_K\geq 2$, so
\[
|K(u)g(x+\sigma u)\mu(x+\sigma u)|\leq C_KC_gC_{\mu}\,n^{-4}e^{-\|u\|^2/(2C_K)},
\]
\[
K_{\sigma}(x,y)\leq \sigma^{-d}C_K \exp(-\frac{r^2}{C_K\sigma^2})=\sigma^{-d}C_Kn^{-4C_K}\leq \sigma^{-d}C_K n^{-8}\quad\text{for } y\in B^c_r(x).
\]
Integrating the first bound over $\|u\|>r/\sigma$ gives at most $C_KC_gC_\mu(2\pi C_K)^{d/2}n^{-4}\leq n^{-3}$; the second claim follows by the change of variables $y=x+\sigma u$, and the third from the second bound.
\end{proof}

\subsection{Proof of Lemma~\ref{lem:ind}}
\begin{proof}
The upper bound is straightforward. For the mean, using Lemma \ref{lem:outlier} we have 
\[
\left|\E[F_i]-\int_{B_r(x_0)} K_\sigma(x_0,y)f(y)\mu(y)dy\right|=O(n^{-3}\sigma^{-d}).
\]
Write $g=f\mu$, which is $\mathcal C^s$ at $x_0$ with constant $L_g$, and set
$M_s:=\int_{\reals^d}|K(u)|\,\|u\|^s\,du<\infty$, finite by $|K(u)|\le C_K e^{-\|u\|^2/C_K}$.
For $n$ large, $\sigma\|u\|\le r\le r_0$ on $\{\|u\|\le r/\sigma\}$, so the Taylor bound applies:
\[
\Big|g(x_0+\sigma u)-g(x_0)-\sum_{j=1}^{s-1}\tfrac{\sigma^j}{j!}\nabla^j g(x_0)[u,\dots,u]\Big|\le L_g(\sigma\|u\|)^s .
\]
Hence
\[
\int_{B_r(x_0)}K_\sigma(x_0,y)g(y)\,dy
=\int_{\|u\|\le r/\sigma}K(u)\Big(g(x_0)+\sum_{j=1}^{s-1}\tfrac{\sigma^j}{j!}\nabla^j g(x_0)[u,\dots,u]\Big)du+R_1,
\]
with $|R_1|\le L_g\sigma^s\int_{\reals^d}|K(u)|\|u\|^s du=L_g M_s\sigma^s=O(\sigma^s)$.
Extending the polynomial integral to $\reals^d$ costs only
\[
|R_2|\le\sum_{j=0}^{s-1}\tfrac{\sigma^j}{j!}\|\nabla^j g(x_0)\|\int_{\|u\|>r/\sigma}|K(u)|\|u\|^j du=O(n^{-3}),
\]
by the same sub-Gaussian tail as Lemma~\ref{lem:outlier} (recall $r/\sigma=2C_K\sqrt{\log n}$).
Finally, by \eqref{eqn:Ksorder}, $\int_{\reals^d}K(u)du=1$ and $\int_{\reals^d}K(u)u^{\otimes j}du=0$ for $1\le j\le s-1$, so the polynomial integral equals $g(x_0)$. Therefore
\[
\int_{B_r(x_0)}K_\sigma(x_0,y)g(y)\,dy=g(x_0)+O(\sigma^s).
\]
Meanwhile, 
\begin{align*}
\text{var}[F_i]\leq \E[F_i^2]&=\int_{\reals^d} K_\sigma(x_0,y) (K_{\sigma}(x_0,y)f(y)^2)\mu(y)dy\\
&=\int_{B_r(x_0)} K_\sigma(x_0,y) (K_{\sigma}(x_0,y)f(y)^2)\mu(y)dy+O(n^{-3}\sigma^{-2d})\\
&=O(V_dr^d\sigma^{-2d}+n^{-3}\sigma^{-2d}). 
\end{align*}

\end{proof}

\subsection{Proof of Proposition~\ref{pro:NW}}
\begin{proof}
Note that by Lemma \ref{lem:ind}, if we write $U_i=F_i-\E[F_i]$, then by Bernstein inequality, 
\[
P\left(|\sum_{i\in [n]} U_i|>t\right)\leq 2\exp\left(-\frac{t^2}{C nr^d \sigma^{-2d}+Ct\sigma^{-d}}\right)
\]
So if we pick $t=C_1 \log n \sqrt{nr^d}\sigma^{-d}$ for some sufficiently large $C_1$, we have
$P(|\sum_{i\in [n]} U_i|>t)\leq n^{-5}$. Indeed, the exponent is at least
\[
\min\big\{\frac{t^2}{2Cnr^d\sigma^{-2d}},\frac{t}{2C\sigma^{-d}}\big\}
=\min\big\{\frac{C_1^2(\log n)^2}{2C},\frac{C_1\log n\sqrt{nr^d}}{2C}\big\},
\]
and
$nr^d=(2C_K)^d n\sigma^d(\log n)^{d/2}\geq (2C_K)^d$ since $n\sigma^d\geq 1$. Meanwhile, Lemma \ref{lem:ind} indicates
\[
|\E[F_i]-f(x_0)\mu(x_0)|=O(\sigma^s).
\]
Putting both parts together we have our claim. 
\end{proof}

\subsection{Proof of Lemma~\ref{lem:buildK}}
\begin{proof}
Note that $\int_{\reals^d} K(\|u\|)du=1$ if $\sum_i a_i=1$. The requirement $\int_{\reals^d}K(\|u\|)\,u^{\otimes j}du=0$ holds for all odd $j$ by symmetry. If $j$ is even, then $\int_{\reals^d}u^{\otimes j}\phi_\sigma(\|u\|)du=\sigma^j\int_{\reals^d}u^{\otimes j}\phi_1(\|u\|)du$, where the last integral is a nonzero tensor. So the requirement is equivalent to
$\sum_{i=1}^{s} (i)^ja_i=0$. So it suffices for us to solve $s$ linear equations
\[
\sum_{i=1}^s a_i=1,\quad \sum_{i=1}^{s} (i)^{j}a_i=0,\quad j=1,\ldots,s-1. 
\]
Note that the coefficient of these equations form a Vandermonde matrix, which will not be singular. Therefore there will always be a unique solution to our problem. Finally, because $K$ is a linear combination of Gaussian density, so $|K(u)|\leq C_K\exp(-|u|^2/C_K)$ holds with some $C_K\geq 2$.    
\end{proof}

\end{document}